\documentclass[12pt]{iopart}
\usepackage{iopams}  
\usepackage{graphicx}
\usepackage{color}
\usepackage{bm}
\usepackage[colorlinks=true,breaklinks=true,allcolors=blue]{hyperref}

\begin{document}
\title[Entanglement Hamiltonian of the 1+1-D free, compactified boson CFT]{Entanglement Hamiltonian of the 1+1-dimensional free, compactified boson conformal field theory}
\author{Ananda Roy}
\ead{ananda.roy@tum.de}
\address{Department of Physics, T42, Technische Universit\"at M\"unchen, 85748 Garching, Germany}
\author{Frank Pollmann}
\address{Department of Physics, T42, Technische Universit\"at M\"unchen, 85748 Garching, Germany}
\address{Munich Center for Quantum Science and Technology (MCQST), 80799 Munich, Germany}
\author{Hubert Saleur}
\address{Institut de Physique Th\'eorique, Paris Saclay University, CEA, CNRS, F-91191 Gif-sur-Yvette.}
\address{Department of Physics and Astronomy, University of Southern California, Los Angeles, CA 90089-0484, USA}

\vspace{10pt}

\begin{abstract}
Entanglement or modular Hamiltonians play a crucial role in the investigation of correlations in quantum field theories. In particular, in 1+1 space-time dimensions, the spectra of entanglement Hamiltonians of conformal field theories (CFTs) for certain geometries are related to the spectra of the physical Hamiltonians of corresponding boundary CFTs. 
As a result, conformal invariance allows exact computation of the spectra of the entanglement Hamiltonians for these models. In this work, we perform this computation of the spectrum of the entanglement Hamiltonian for the free compactified boson CFT over a finite spatial interval. 
We compare the analytical results obtained for the continuum theory with numerical simulations of a lattice-regularized model for the CFT using density matrix renormalization group technique. To that end, we use a lattice regularization provided by superconducting quantum electronic circuits, built out of Josephson junctions and capacitors. 
Up to non-universal effects arising due to the lattice regularization, the numerical results are compatible with the predictions of the exact computations. 
\end{abstract}
\maketitle 

\section{Introduction}
Entanglement plays an indispensable role in the analysis of correlations present in quantum field theories. The von-Neumann entanglement entropy, $S(\rho_A) = -{\rm Tr}\rho_A\ln\rho_A$, is one of the most popular measures of bipartite entanglement~\cite{Eisert2010}. Here, $\rho_A$ is the reduced density matrix of the subsystem A: $\rho_A = {\rm Tr}_B\rho$, where $\rho$ is the total density matrix of the system composed of parts A and B. 
%In two spatial dimensions, for systems exhibiting topological order, the entanglement entropy of the ground state at zero temperature  contains signatures of the quasiparticle  content of the theory~\cite{Hamma2005, Kitaev2006a, Levin2006}. Thus, it can be used to characterize the anyonic content of the topological field theory describing the topological phase. 
The entanglement entropy is crucial in the characterization of quantum field theories in 1+1 space-time dimensions. The scaling of the entanglement entropy in critical systems in 1+1 dimensions has been predicted using conformal field theory (CFT) techniques~\cite{Holzhey1994, Calabrese2004, Korepin2004}. These have been used to describe the quantum critical phenomena in 1D spin-chains~\cite{Vidal2003, Pollmann2009} as well as observe boundary-RG flow between different conformal invariant boundary conditions~\cite{Affleck1991, Affleck2001, Affleck2009}.

\begin{figure}
 \centering
\includegraphics[width = 0.7\textwidth]{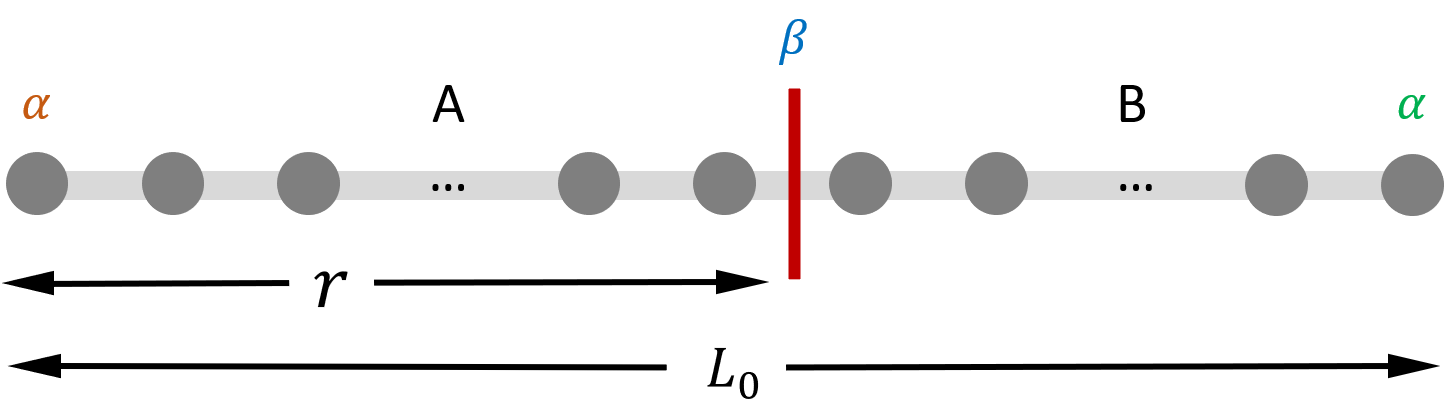}
\caption{\label{1D_system} Schematic of a 1D lattice model of length $L_0$ bipartitioned into subsystems A (length $r$) and B (length $L_0-r$). The von-Neumann entanglement entropy for the subsystem A is given by $S(\rho_A) = -{\rm Tr}\rho_A\ln\rho_A$, where $\rho_A$ is the reduced density matrix of the subsystem A. The entanglement Hamiltonian is defined as ${\cal H}_A = -(\ln\rho_A)/2\pi$ and is not, in general, the Hamiltonian of the subsystem A. Here, $\alpha$ denotes the boundary conditions of the original system and the boundary condition $\beta$ originates from the entanglement cut. }
\end{figure}

%One of the most appealing feature of von-Neumann entropy as an entanglement measure is its ease of computation -- a single number has to be computed. 
In 1+1 dimensional CFTs, the scaling of the entanglement entropy with the size of the subsystem A enables the determination of the central charge or the conformal anomaly parameter without the need to determine the velocity of sound in the theory~\cite{Calabrese2004}. However, the full operator content of the CFT remains elusive. The latter can be probed through the spectrum of the entanglement or modular Hamiltonian of the subsystem A (see Fig.~\ref{1D_system}) defined as~\cite{Li2008, Haag2012}
\begin{equation}
\label{ent_ham}
{\cal H}_A =-\frac{1}{2\pi}\ln\rho_A,
\end{equation}
where we follow the convention of Ref.~\cite{Cardy2016}. 
It turns out that the spectrum of the entanglement Hamiltonian of a CFT is given by the physical spectrum of a corresponding boundary CFT~\cite{Cardy2016, DiGiulio2019, DeLuca2013}. The relationship of the entanglement Hamiltonian to the physical Hamiltonian of a boundary CFT opens the possibility to determine exactly the spectrum of the entanglement Hamiltonian. This can be done by computing boundary/Ishibashi states of the theory and subsequently, the partition function of the boundary CFT (see Chap. 11 of Ref.~\cite{diFrancesco1997} or Ref.~\cite{Cardy1989} for details of the formalism).
The main goal of this work is perform this computation for the free, compactified boson CFT and provide numerical data obtained with density matrix renormalization group (DMRG) technique for a lattice-regularized model. We note that DMRG results were obtained earlier for the critical transverse-field Ising chain and the Bose-Hubbard model in Ref.~\cite{Lauchli2013} (see also Refs.~\cite{Giudici2018, Zhang2020}), which had suggested the boundary CFT structure of the entanglement Hamiltonian. In this work, we go a step further and perform analytical computations for the different boundary conditions and conduct a careful comparison of the DMRG results for finite system sizes.

We focus on the case when the system under investigation is finite (with length $L_0$) with a certain prescribed boundary condition, $\alpha$ at its ends. We treat only the case of identical boundary conditions at both ends (the case with different boundary conditions at the two ends has additional technical complications~\cite{Cardy2016}, which we leave for a later work) for a system at zero temperature and consider a subsystem A of length $r$. For this case, the spectrum of ${\cal H}_A$ is determined by that of the Hamiltonian $H_{\alpha\beta}$ of the boundary CFT with boundary conditions $\alpha,\beta$, where $\alpha\neq\beta$ in general~\cite{Cardy2016}. The first boundary condition $\alpha$ is inherited from the original system, while the second $\beta$ originates from the entanglement cut and is usually the free boundary condition. Thus, for the case when $\alpha$ corresponds to free boundary conditions, the boundary CFT also has free boundary conditions at both ends. On the other hand, if $\alpha$ corresponds to fixed boundary conditions, then the corresponding boundary CFT has fixed and free boundary conditions at its ends. The final result for the entanglement Hamiltonian is given by~\cite{Cardy2016}
\begin{equation}
\label{ent_ham_formula}
{\cal H}_A = -\frac{1}{2\pi}\ln \frac{e^{-2\pi H_{\alpha\beta}}}{{\rm Tr}\ e^{-2\pi H_{\alpha\beta}}},
\end{equation}
where the denominator inside the logarithm originates from the fact that the reduced density matrix $\rho_A$ should be normalized. The above equation is to be understood as an equality of the eigenvalues of the two sides the equation up to overall shifts and rescalings, which can be absorbed by rescaling the velocity of sound in the corresponding boundary CFT. 

To illustrate the basic principle of the analysis, we start by deriving the entanglement Hamiltonian of the Ising CFT using the exact results for the Ishibashi states obtained by Cardy~\cite{Cardy1989} and compare with DMRG results of the corresponding lattice model of the critical transverse-field Ising chain. Then, we present the exact results for the free compactified boson CFT. Generalizing the computation done in Ref.~\cite{Oshikawa1997}, we provide an explicit closed form expression for the spectrum of the entanglement Hamiltonian in terms of the compactification radius and the system size. In order to compare our analytical predictions of the continuum theory for a lattice model, we analyze a lattice regularized model using quantum electronic circuits~\cite{Bradley1984, Korshunov1989, Glazman1997, Goldstein2013, Roy2019} using DMRG. This should be contrasted with usual lattice-regularizations of the free, compactified boson CFT using the paramagnetic phase of the XXZ spin chain~\cite{Giamarchi2003, Gogolin2004}.~\footnote{We focus on the compact boson case. The non-compact case can be treated by considering massless harmonic oscillator chains~\cite{DiGiulio2019, Peschel1991}.} In contrast to the latter model, we start from compact, bosonic lattice degrees of freedom. The quantum circuit is a 1D array of superconducting islands, separated by tunnel junctions, realizing a generalized Bose-Hubbard model in the limit of high-occupancy of each site~\cite{Roy2020}. Each superconducting island has a finite charging energy $E_{C_0} = 2e^2/C_0$, where $C_0$ is the capacitance to the ground plane. The Josephson junction separating two such islands has a junction energy $E_J$ and charging energy $E_{C_J} = 2e^2/C_J$. The lattice model has a rich phase-diagram comprising Mott-insulating, charge-density-wave and the free, compactified boson phases. We focus on the latter phase, which occurs in the regime $E_{C_J}, E_{C_0} \ll E_J$. The compactification radius $R = 1/\beta = 1/\sqrt{\pi K}$, where $K$ is the Luttinger parameter of the system. The compactified bosonic field $\phi(x,t)$ is the Josephson phase on the island at position $x$ at time $t$. We analyze the entanglement Hamiltonian of the CFT under consideration using DMRG. 

The article is organized as follows. In Sec.~\ref{IsingCFT}, we derive the spectrum of the entanglement Hamiltonian of the Ising CFT and compare with that obtained using DMRG. Subsequently in Sec.~\ref{free_boson_CFT}, we present our results for the free, compactified boson CFT. The analytical, exact results are presented in Sec.~\ref{ent_ham_exact}. The details of the computation of the boundary states are provided in~\ref{bcft}, where we compute the partition function of the compactified boson CFT in the presence of different boundary conditions. In Sec.~\ref{dmrg}, we provide DMRG analysis of free, compactified boson phase of the lattice quantum circuit model. The detailed phase-diagram of the lattice model, while interesting on its own, is not relevant for the main goal of the work, and is given in~\ref{phase_diag}. In Sec.~\ref{concl}, we summarize our findings and provide a concluding perspective. 

\section{A simple test case: the Ising CFT}
\label{IsingCFT}
To illustrate the basic principle of the analysis, in this section, we derive the spectrum of the entanglement Hamiltonian for the Ising CFT. The latter is the unitary, minimal model ${\cal M}(4,3)$ with central charge $c = 1/2$ (see, for example, Chapters 7 and 8 of Ref.~\cite{diFrancesco1997}). It contains three primary fields, $I, \sigma, \epsilon$, with conformal dimensions: $h_{0}=0, h_{\sigma}=1/16$ and $h_{\epsilon}=1/2$.~\footnote{We have used $\epsilon$ with a subscript to denote the eigenvalues of the Hamiltonian of the corresponding boundary CFTs in~\ref{bcft}, but there should not be any confusion. } 
\subsection{Exact results}
We will consider two cases: (i) free/Neumann (N) boundary conditions at both ends ($\alpha={\rm N}$) and (ii) fixed/Dirichlet (D) boundary conditions at both ends ($\alpha={\rm D}$). Thus, the entanglement Hamiltonian for the subsystem $A$ (see Fig.~\ref{1D_system}) will be given by the spectrum of the boundary CFT over a length $L$ with boundary conditions $\alpha=\beta={\rm N}$ in the first case and $\alpha={\rm D}$, $\beta={\rm N}$ in the latter. We emphasize that $L$ is {\it not} the length of the subsystem A, the latter being denoted by $r$. The relation between $L_0, L$ and $r$, to leading order, is given by
\begin{equation}
\label{ldef}
L = \ln\Big(\frac{2L_0}{\pi a}\sin\frac{\pi r}{L_0}\Big),
\end{equation}
with correction arising at ${\cal O}(a)$. Here, $a$ is the lattice spacing. The partition function for the boundary CFTs can be expressed in terms of the parameters $q, \tilde{q}$:
\begin{equation}
\label{qdef}
q = e^{-2\pi^2/L}, \ \tilde{q} = e^{-2L}.
\end{equation}
Note that as $L\rightarrow \infty$, $q\rightarrow1$ and $\tilde{q}\rightarrow0$. So, it is more convenient to express final results as series in $\tilde{q}$ rather than $q$ for better convergence. The boundary states for the different boundary conditions are given by~\cite{Cardy1989}
\begin{eqnarray}
|\tilde{0}\rangle &= \frac{1}{\sqrt{2}}|0\rangle + \frac{1}{\sqrt{2}}|\epsilon\rangle + \frac{1}{2^{1/4}}|\sigma\rangle,\\
\Big|\tilde{\frac{1}{2}}\Big\rangle &= \frac{1}{\sqrt{2}}|0\rangle + \frac{1}{\sqrt{2}}|\epsilon\rangle - \frac{1}{2^{1/4}}|\sigma\rangle,\\ 
\Big|\tilde{\frac{1}{16}}\Big\rangle &= |0\rangle - |\epsilon\rangle,
\end{eqnarray}
where the first two correspond to Dirichlet boundary conditions and the last corresponds to Neumann boundary condition.  
For case (i), the corresponding partition function can be written as a sum over characters of the Ising CFT:
\begin{eqnarray}
Z_{\rm{NN}}(q) &= {\rm Tr}e^{-2\pi H_{\rm NN}}= \sum_{j=0,\sigma,\epsilon}\Big|\Big\langle \tilde{\frac{1}{16}}\Big|j\Big\rangle\Big|^2 \chi_j(\tilde{q})\nonumber\\& = \chi_0(\tilde{q}) + \chi_{\epsilon}(\tilde{q})\\&=\chi_0(q) + \chi_{\epsilon}(q),
\end{eqnarray}
where in the last line, we have used the explicit form of the modular S-matrix of the Ising CFT~\cite{Cardy1989}. Thus, we find that the partition function gets contribution from two primary fields: $I,\epsilon$. We use the explicit formulas for the characters (see Chapter 8 of Ref.~\cite{diFrancesco1997}):
\begin{eqnarray}
\chi_0(q) &= \frac{1}{\eta(q)}\sum_{n\in\mathbb{Z}}\Big[q^{(24n+1)^2/48} - q^{(24n+7)^2/48}\Big],\\\chi_{\epsilon}(q) &= \frac{1}{\eta(q)}\sum_{n\in\mathbb{Z}}\Big[q^{(24n+5)^2/48} - q^{(24n+11)^2/48}\Big],
\end{eqnarray}
where $\eta(q)$ is the Dedekind function defined as 
\begin{equation}
\label{Dedekind}
\eta(q) = q^{1/24}\varphi(q) = q^{1/24}\prod_{n>0} (1-q^n). 
\end{equation}
Expanding in $q$, we get 
\begin{eqnarray}
\chi_j(q) &=q^{-1/48+h_j}\sum_{n\geq0}p_j(n)q^n,\ j = 0,\epsilon,
\end{eqnarray}
where $p_{0,\epsilon}(i)$ are obtained to be
\begin{eqnarray}
p_0(n) &= 1,0,1,1,2,2,3,\ldots,\\ p_\epsilon(n)&=1,1, 1, 1, 2, 2, 3,\ldots.
\end{eqnarray}
Thus, the entanglement energies, labeled by two indices: $(j,n)$, are given by
\begin{eqnarray}
  \label{eNN}
\varepsilon_{\rm{N}}(j,n) &= -\frac{1}{2\pi}\ln \frac{q^{-1/48 + h_j+n}}{\tilde{q}^{-1/48}\sum\limits_{k=0,\epsilon}\sum\limits_{m\geq0}p_k(m){\tilde{q}}^{h_k+m}}\nonumber\\&=\frac{L}{48\pi} +\frac{\pi}{L}\Big(-\frac{1}{48}+h_j + n\Big)\nonumber\\&\quad +\frac{1}{2\pi}\ln\sum\limits_{k=0,\epsilon}\sum\limits_{m\geq0}p_k(m)e^{-2L(h_k+m)}
\end{eqnarray}
with degeneracy at the level $(j,n)$ being given by $p_j(n)$. The lowest entanglement energy level is given by
\begin{eqnarray}
\label{e0_tfi_NN}
\varepsilon_{\rm{N}}(0,0) &= \frac{L}{48\pi} -\frac{\pi}{48L}+\frac{1}{2\pi}\ln\sum\limits_{k=0,\epsilon}\sum\limits_{m\geq0}p_k(m)e^{-2L(h_k+m)}.
\end{eqnarray}
With respect to this lowest level, the entanglement energies are given by 
\begin{equation}
\Delta\varepsilon_{\rm{N}}(j,n) \equiv \varepsilon_{\rm{N}}(j,n) - \varepsilon_{\rm{N}}(0,0) = \frac{\pi}{L}\big(h_j + n\big),
\end{equation}
and thus, occur at integer (half-integer) values in units of $\pi/L$ for $j=0(\epsilon)$. 

For case (ii), the analysis proceeds analogously. The partition function for the corresponding boundary CFT with Dirichlet and Neumann boundary conditions at the ends is
\begin{eqnarray}
Z_{\rm{DN}}(q) &= \chi_\sigma(q) = \frac{1}{\sqrt{2}}[\chi_0(\tilde{q}) - \chi_\epsilon(\tilde{q})].
\end{eqnarray}
Here, \begin{eqnarray}
\chi_\sigma(q)&= \frac{1}{\eta(q)}\sum_{n\in\mathbb{Z}}\Big[q^{(24n-2)^2/48} - q^{(24n+10)^2/48}\Big]\nonumber\\&=q^{-1/48+h_\sigma}\sum_{n\geq0}p_\sigma(n)q^n,
\end{eqnarray}
where we obtain
\begin{equation}
\label{psigma}
p_\sigma(n) = 1,1,1,2,2,3,\ldots.
\end{equation}
In this case, the entanglement energies, indexed by $n$, are given by 
\begin{eqnarray}
\label{e_tfi_DN}
\varepsilon_{\rm{D}}(n) &= -\frac{1}{2\pi}\ln \frac{\sqrt{2}q^{-1/48 + h_\sigma+n}}{\tilde{q}^{-1/48}\sum\limits_{k=0,\epsilon}e^{2\pi ih_k}\sum\limits_{m\geq0}p_k(m){\tilde{q}}^{h_k+m}}\nonumber\\&=\frac{L}{48\pi}-\frac{1}{4\pi}\ln2 +\frac{\pi}{L}\Big(\frac{1}{24} + n\Big)\nonumber\\&\quad +\frac{1}{2\pi}\ln\sum\limits_{k=0,\epsilon}e^{2\pi ih_k}\sum\limits_{m\geq0}p_k(m)e^{-2L(h_k+m)},
\end{eqnarray}
where the degeneracy at level $n$ is given by $p_\sigma(n)$. The lowest energy level $\varepsilon(0)$ is given by 
\begin{eqnarray}
\label{e0_tfi_DN}
\varepsilon_{\rm D}(0) &= \frac{L}{48\pi}-\frac{1}{4\pi}\ln2 +\frac{\pi}{24L}\nonumber\\&\quad +\frac{1}{2\pi}\ln\sum\limits_{k=0,\epsilon}e^{2\pi ih_k}\sum\limits_{m\geq0}p_k(m)e^{-2L(h_k+m)},
\end{eqnarray}
with respect to which the entanglement energies are given by 
\begin{equation}
\Delta\varepsilon_{\rm{D}}(n) \equiv\varepsilon_{\rm{D}}(n) - \varepsilon_{\rm{D}}(0) = \frac{\pi}{L}n
\end{equation}
Recall that the entanglement entropy for the subsystem A is given by~\cite{Calabrese2004}
\begin{equation}
\label{S}
{\cal S}(L) = \frac{c}{6}L + {\cal O}(1),
\end{equation}
where $c=1/2$. The last term contains the boundary terms predicted by Affleck and Ludwig as well as a non-universal correction. Comparing the leading order terms in either of  Eqs.~(\ref{e0_tfi_NN},\ref{e0_tfi_DN}) with Eq.~(\ref{S}), we get the expected relation between the entanglement entropy and the single-copy entanglement~\cite{Eisert2005}
\begin{equation}
\label{single_copy_ent}
\varepsilon_{\rm{D/N}}(0,0) = \frac{1}{4\pi}{\cal S} + {\cal O}(1),
\end{equation}
where we have an extra factor of $2\pi$ due to our definition of Eq.~(\ref{ent_ham}). Furthermore, it is useful to compare the lowest entanglement energies for the two cases obtained in Eqs.~(\ref{e0_tfi_NN}, \ref{e0_tfi_DN}). For $L\rightarrow\infty$, the difference between the two is given by the term of ${\cal O}(1)$ in Eq.~(\ref{e0_tfi_DN}):
\begin{equation}
  \label{ising_bdry}
\varepsilon_{\rm{N}}(0,0) - \varepsilon_{\rm{D}}(0) = \frac{1}{4\pi}\ln 2.
\end{equation}
This difference is the change in the Affleck-Ludwig boundary entropy as we go from Neumann-Neumann to Dirichlet-Neumann boundary conditions in the boundary CFT. Note that the relationship between the single-copy entanglement and the entanglement entropy [Eq.~(\ref{single_copy_ent})] does not hold for the ${\cal O}(1)$ term. Furthermore, there is a difference by a factor of $2\pi$ with the original work~\cite{Affleck1991} due to conventions chosen in Eq.~(\ref{ent_ham}). 

\subsection{DMRG results}
\begin{figure}
\centering
\includegraphics[width = \textwidth]{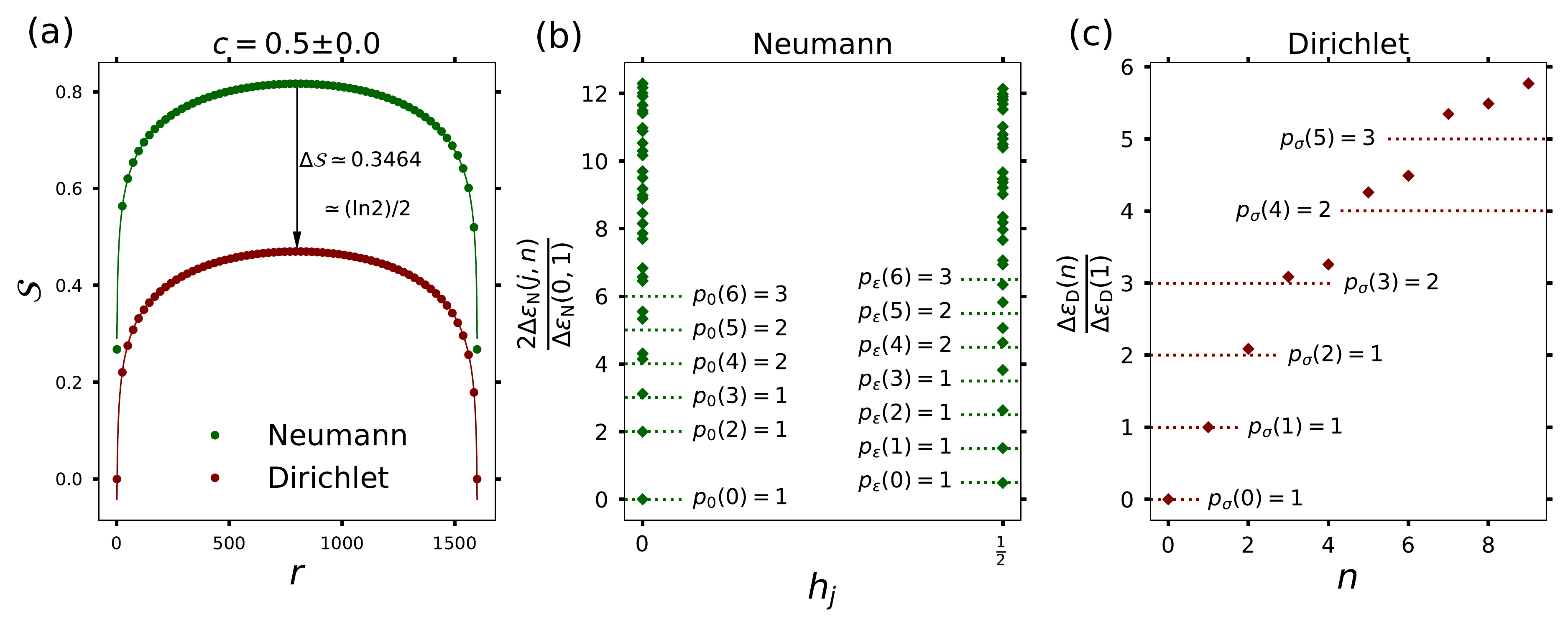}
  \caption{\label{NN_TFI} DMRG results for the critical transverse field Ising chain. The system size $L_0 = 1600$. (a) Entanglement entropy ${\cal S}$ as a function of the subsystem size $r$ for Neumann (maroon) and Dirichlet (dark green) boundary conditions. The central charge was verified to be $\simeq1/2$ by fitting to Eq.~(\ref{ent_form}). The shown value of $c$ is obtained by fitting to the data for the Neumann boundary condition. As the boundary condition changes from Neumann to Dirichlet, the entanglement entropy changes by $0.3464$ which is close to the expected change in the boundary entropy given by $(\ln 2)/2$.
 (b) Renormalized entanglement spectrum by partitioning the system at $L_0/2$ for Neumann boundary condition. The predictions from the CFT are shown for the lowest few levels with dashed lines. We see that the agreement with the CFT predictions are good for the low-lying entanglement energy levels. The agreement worsens for higher energy levels due to the finite-size effects. We checked the latter assertion by varying the system size.
  (c) Renormalized entanglement spectrum by partitioning the system at $L_0/2$ for Dirichlet boundary condition. The agreement with the CFT predictions (shown in dashed lines) is good for the low-lying levels. We note that the finite size effects are larger in this case compared to the Neumann case, but we checked that the results approached the CFT predictions as system sizes were increased.  
  }
\end{figure}

\begin{figure}
\centering
\includegraphics[width = 0.65\textwidth]{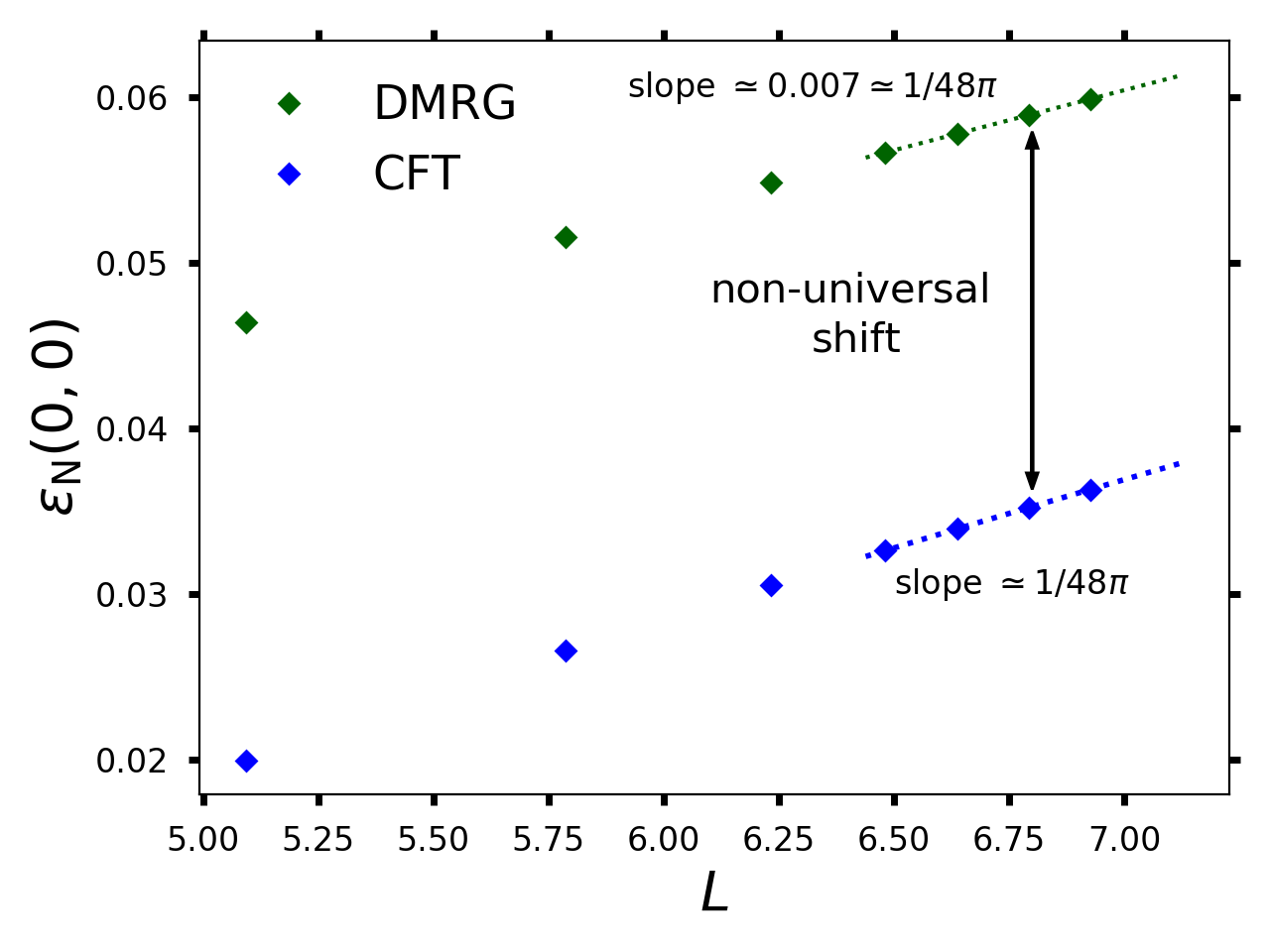}
\caption{\label{e0_TFI} Comparison of the lowest entanglement energy level obtained by DMRG (shown in maroon) with the CFT predictions (shown in blue) for Neumann boundary condition. The DMRG simulations were performed for system sizes $L_0 = 256,$ 512, 800, 1024, 1200, 1400 and 1600. The size of the subsystem A was chosen to be $r=L_0/2$, which leads to $L = \ln(2L_0/\pi)$ [see Eq.~\ref{ldef}].	We see that up to an overall shift of the energy levels that tends to a constant value of $\simeq0.24$, the DMRG results are consistent with the CFT predictions. }
\end{figure}

In this section, we compute using DMRG, the entanglement spectrum of the critical transverse-field Ising chain and compare with the analytical CFT predictions derived above. We show the DMRG results for the cases when either Neumann or Dirichlet boundary conditions are imposed on both ends of the chain [referred to as cases (i) or (ii) above]. We chose the system size to be $L_0=1600$ and a bond-dimension of $600$ to keep truncation errors below $10^{-12}$. We verify the central charge ($c$) to be $\simeq1/2$. This is done by evaluating the entanglement entropy ${\cal S}$ for a finite block (of length $r$) within the system (of length $L_0$), using Eqs.~(\ref{ldef}, \ref{S}). Explicitly, 
\begin{equation}
  \label{ent_form}
{\cal S}(r,L_0) = \frac{c}{6}\ln \Big(\frac{2L_0}{\pi a}\sin\frac{\pi r }{L_0}\Big) + {\cal S}_0,
\end{equation}
where ${\cal S}_0$ contains the contribution from the boundary as well as non-universal terms. By changing the boundary conditions, we obtain a change in entropy that is very close to the expected value of $(\ln2)/2$ [see Fig.~\ref{NN_TFI}~(a)] (note that the change in entropy is $2\pi$ times the change in the lowest entanglement energy in our convention).  
In Fig.~\ref{NN_TFI}(b), we show the rescaled and shifted entanglenent spectrum obtained for Neumann boundary conditions for a partitioning at the center of the center of the chain ($r=L_0/2$). As predicted by Eq.~(\ref{eNN}), the spectrum is indeed split into two Virasoro blocks, corresponding to the primary fields $0$ and $\epsilon$. The rescaled spectrum is close to the CFT predictions indicated by the dashed lines. 
In  Fig.~\ref{NN_TFI}(c), we show the same for Dirichlet boundary conditions partitioning the chain in two halves. From Eq.~(\ref{e_tfi_DN}), there is only one Virasoro block corresponding to the primary field $\sigma$. The obtained entanglement spectrum is close to the CFT predictions (shown in dashed lines). The finite size effects are larger in this case, compared to the Neumann case, but we verified that the discrepancy between the DMRG results and the CFT predictions diminish upon increasing the system size. 
Next, we compare the DMRG results for the lowest entanglement spectrum eigenvalue with that obtained from CFT (Fig.~\ref{e0_TFI}). We do this for the Neumann case, similar results were obtained for the Dirichlet case. Up to an overall shift, which for large system sizes is a constant $\simeq0.24$, the DMRG results exhibit the same asymptotic behavior as system sizes are increased [note that the variation is plotted with respect to $L = \ln(2L_0/\pi)$]. 

\section{The free compactified boson CFT}
\label{free_boson_CFT}
In this section, we compute the spectrum of the entanglement Hamiltonian of the free compactified boson CFT. The compactification radius is given by $R=1/\sqrt{\pi K}$, where $K$ is the Luttinger parameter of the theory. The analytical results were obtained by extending the calculations of Ref.~\cite{Oshikawa1997}, for more details see~\ref{bcft}. 

\subsection{Exact results}
\label{ent_ham_exact}
We consider a finite system of size $L_0$ and analyze the cases when the system has (i) free/Neumann boundary conditions and (ii) fixed/Dirichlet boundary conditions at both ends. Thus, the spectrum of the boundary CFT that needs to be evaluated has Neumann-Neumann and Dirichlet-Neumann boundary conditions at two ends of the interval of length $L$ for the two cases respectively. These can be obtained from the corresponding expressions in~\ref{bcft} by setting $T = 2\pi$. 

Consider case (i). In terms of these two parameters $q, \tilde{q}$, the corresponding partition function for the boundary CFT is given by 
\begin{eqnarray}
Z_{\rm{NN}} &= {\rm Tr} \ e^{-2\pi H_{\rm{NN}}}\\\label{q_expr}&= \sum_{k\in\mathbb{Z}}\sum_{l\geq0}p(l)q^{-\frac{1}{24} +\frac{K}{2}k^2 + l}\\\label{qt_expr}&=\frac{1}{\sqrt{K}\eta(\tilde{q})}\sum_m \tilde{q}^{\frac{m^2}{2K}}\\&=\sum_h n^h_{\rm{NN}}\chi_h(q),
\end{eqnarray}
where $\eta(q)$ is the Dedekind function and and $p(l)$ is the number of integer partitions of the integer $l$. In the last line, we have expressed the partition function as a sum over the Virasoro characters for different primary fields with dimension $h$~\cite{Cardy1989}(see~\ref{bcft} for more details). Here, we have also used the fact that the central charge for the CFT is 1. The expression in Eq.~(\ref{q_expr}) shows that the spectrum is composed of Virasoro towers built on primary fields with dimension $Kk^2/2$, $k\in\mathbb{Z}$. The towers on top of each primary are themselves built out of the descendants indexed by $l$ and each level has degeneracy $p(l)$. Denote the eigenvalues of the entanglement Hamiltonian ${\cal H}_A$ by $\varepsilon_{\rm{N}}(k,l)$ where it is implied that the degeneracy for each $l$ is $p(l)$. Then, 
\begin{eqnarray}
\varepsilon_{\rm{N}}(k,l) &= -\frac{1}{2\pi}\ln \frac{q^{-\frac{1}{24} +\frac{K}{2}k^2 + l}}{\frac{1}{\sqrt{K}\eta(\tilde{q})}\sum_{m\in\mathbb{Z}} \tilde{q}^{\frac{m^2}{2K}}}\nonumber
%\\&=-\frac{1}{4\pi}\ln K -\frac{\ln q}{2\pi}\Big(-\frac{1}{24} +\frac{K}{2}k^2 + l\Big)\nonumber\\&\quad-\frac{1}{2\pi}\ln\eta(\tilde{q})+\frac{1}{2\pi}\ln\Big(\sum_m \tilde{q}^{\frac{m^2}{2K}}\Big)
\\&= \frac{L}{24\pi}-\frac{1}{4\pi}\ln K +\frac{\pi}{L}\Big(-\frac{1}{24} +\frac{K}{2}k^2 + l\Big)\nonumber\\&\quad -\frac{1}{2\pi}\Bigg[\sum_{n>0}\ln(1-e^{-2Ln})-\ln\sum_{m\in\mathbb{Z}} e^{-\frac{Lm^2}{K}}\Bigg]. 
\end{eqnarray}
The smallest eigenvalue $\varepsilon_{\rm{N}}(0,0)$ is given by
\begin{eqnarray}
\label{e_0}
\varepsilon_{\rm{N}}(0,0) &= \frac{L}{24\pi}-\frac{1}{4\pi}\ln K -\frac{\pi}{24L}\nonumber\\& -\frac{1}{2\pi}\Bigg[\sum_{n>0}\ln(1-e^{-2Ln}) -\ln\sum_{m\in\mathbb{Z}} e^{-\frac{Lm^2}{K}}\Bigg]. 
\end{eqnarray}
The last two terms in the expression for $\varepsilon_{\rm N}(0,0)$ contribute to the `unusual' corrections to the entanglement entropy obtained in Ref.~\cite{Cardy2010}. The scaling of the higher entanglement energies is given by
\begin{equation}
\label{NN_disp}
\Delta\varepsilon_{\rm{N}}(k,l)\equiv\varepsilon_{\rm{N}}(k,l) - \varepsilon_{\rm{N}}(0,0) = \frac{\pi}{L}\Big(\frac{K}{2}k^2 + l\Big).
\end{equation}
Now consider case (ii). The relevant partition function [see Eq.~(\ref{part_fun_ND})] is given by 
\begin{eqnarray}
Z_{\rm DN}(q) & = &\frac{1}{\sqrt{2}\eta(\tilde{q})}\sum_n  (-1)^n \tilde{q}^{n^2}= q^{1/48}\sum_{n\geq0}q^{n/2}p_\sigma(n)
\end{eqnarray}
where the $p_\sigma(n)$ is defined in Eq.~(\ref{psigma}). 
Using Eq.~(\ref{ent_ham_formula}), the spectrum of the entanglement Hamiltonian is given by 
\begin{eqnarray}
\varepsilon_{\rm{D}}(n) &= -\frac{1}{2\pi}\ln \frac{\sqrt{2}\eta(\tilde{q})q^{\frac{1}{48} + n/2}}{\sum\limits_{n\in\mathbb{Z}}(-1)^n\tilde{q}^{n^2}}\nonumber\\&= \frac{L}{24\pi}-\frac{1}{4\pi}\ln2 + \frac{\pi}{L}\Big(\frac{1}{48}+ \frac{n}{2}\Big) \nonumber\\&\quad -\frac{1}{2\pi}\Bigg[\sum_{n>0}\ln(1-e^{-2Ln})-\ln\sum_{m\in\mathbb{Z}} (-1)^me^{-2Lm^2}\Bigg],
\end{eqnarray}
where we used the definitions in Eq.~(\ref{qdef}) and $n\geq0$. The degeneracy at level $n$ is given by $p_\sigma(n)$. The smallest eigenvalue is given by 
\begin{eqnarray}
\varepsilon_{\rm{D}}(0) &= \frac{L}{24\pi} - \frac{1}{4\pi}\ln2 + \frac{\pi}{48L} -\frac{1}{2\pi}\Bigg[\sum_{n>0}\ln(1-e^{-2Ln})\nonumber\\&\quad-\ln\sum_{m\in\mathbb{Z}} (-1)^me^{-2Lm^2}\Bigg],
\end{eqnarray}
and the higher entanglement energies are given by 
\begin{eqnarray}
  \label{DN_disp}
\Delta\varepsilon_{\rm{D}}(n)&\equiv\varepsilon_{\rm{D}}(n) - \varepsilon_{\rm{D}}(0) = \frac{\pi}{2L}n.
\end{eqnarray}
We note again the corresponding relation between the single-copy entanglement and the entanglement-entropy as obtained for the Ising CFT in Eq.~(\ref{single_copy_ent}). Furthermore, by comparing the lowest entanglement energies for the two cases, in the limit of $L\rightarrow \infty$, we find
\begin{equation}
  \label{eND_b}
\varepsilon_{\rm{N}}(0,0) - \varepsilon_{\rm{D}}(0) = -\frac{1}{4\pi}\ln\frac{K}{2}, 
\end{equation}
which is the change in the boundary entropy in this case~\cite{Fendley1994}. Fermionizing the compactified boson action in the presence of boundary fields at $K=1$ results in the same change in the boundary entropy as in the Ising case [see Eq.~(\ref{ising_bdry})], a reflection of the well-known correspondence of this model (with boundary fields) with two uncoupled Ising chains with boundary magnetic field on one of them~\cite{Fendley1994}.
 
So far, we have computed exactly the spectrum of the entanglement Hamiltonian. Next, we provide a numerical test for our computations by performing DMRG calculations on a suitably regularized lattice model. 

\subsection{DMRG analysis of the quantum circuit model}
\label{dmrg}
\subsubsection{Description of the model:}
\label{lattice_model}
\begin{figure}
\centering
\includegraphics[width = 0.6\textwidth]{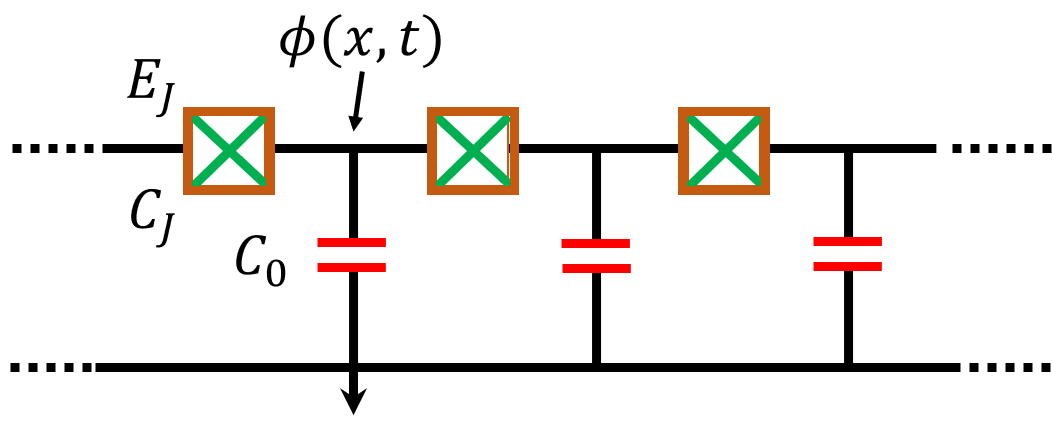}
\caption{\label{free_boson} Schematic of the 1D lattice-regularized model for the free, compactified boson CFT. Each unit cell contains a capacitor (with capacitance $C_0$) on the vertical link and a Josephson junction (with junction energy $E_J$ and junction capacitance $C_J$). In the limit: $E_{C_J}\ll E_{C_0}\ll E_J$, the ground state of this array is described by a free, compactified boson CFT, where the bosonic field is the superconducting phase $\phi(x,t)$ at the indicated node of the lattice. Here, $E_{C_{0,J}} = 2e^2/C_{0,J}$. The compactification radius is given by $R = 1/\sqrt{\pi K}$, where $K$ is the Luttinger parameter.}
\end{figure}

The lattice model for the free-compactified boson CFT comprises a 1D array of mesoscopic superconducting islands separated by tunnel junctions (see Fig.~\ref{free_boson}). Each unit cell contains a capacitor (with capacitance $C_0$) on the vertical link and a Josephson junction (with junction energy $E_J$ and junction capacitance $C_J$). Throughout this work, we assume the absence of disorder in the model. Here, we choose the parameters such that $E_{C_J}\ll E_{C_0}\ll E_J$, where $E_{C_{0,J}} = 2e^2/C_{0,J}$. In this limit, the phase-slips across the array are exponentially suppressed by a WKB factor $\sim e^{-\sqrt{E_J/E_{C_J}}}$. This leads to the low-energy properties of the theory being described by a Luttinger liquid or equivalently a free, compactified boson CFT~\cite{Glazman1997, Goldstein2013, Roy2019}. Here, the superconducting phase, $\phi(x,t)$, at node $(x,t)$ is bosonic field under consideration. 
The effective euclidean action for the theory describing the low-energy physics is given by~\cite{Giamarchi2003}
\begin{eqnarray}
S_{\rm{array}} = \frac{1}{2\pi K}\int dt\int_0^L dx \Big[\frac{1}{u}(\partial_t \phi)^2 + u (\partial_x\phi)^2\Big],
\end{eqnarray}
Here, the plasmon velocity, $u$, and the Luttinger parameter, $K$, are given by~\cite{Goldstein2013}
\begin{equation}
u \simeq a\sqrt{2E_{C_0}E_J},\ K \simeq \frac{1}{2\pi}\sqrt{\frac{2E_{C_0}}{E_J}},
\end{equation}
where $a$ is the lattice spacing. We note that these analytical expressions are only asymptotically true since the lattice model, to the best of our knowledge, is not exactly solvable. In our work, we extract the relevant properties of the model using DMRG. This is done by computing the ground state properties of the following lattice Hamiltonian:
\begin{eqnarray}
\label{ham_arr}
H_{\rm{array}} &= E_{C_0}\sum_{i=1}^Ln_i^2 + \delta E_{C_0}\sum_{i=1}^{L-1}n_in_{i+1} - E_J\sum_{i=1}^{L-1}\cos(\phi_i - \phi_{i+1}).
\end{eqnarray}
Here, the first term arises due to the finite charging energy of the mesoscopic islands and $n_i$ is the excess number of Cooper pairs on the $i^{\rm{th}}$ island~\footnote{Note that $n_i$ can be both positive or negative, the latter corresponding to removal of a Cooper-pair from the superconducting condensate on the $i^{\rm th}$ island.}. The finite junction capacitance $C_J$ leads to, in principle, infinite-range interaction between any two islands with a magnitude that decays exponentially with distance~\cite{Goldstein2013}. However, for realistic system parameters~\cite{Manucharyan2009}, it suffices to include only the nearest neighbor interaction~\cite{Glazman1997}, indicated by the second term in Eq.~(\ref{ham_arr}) with $\delta$ being a small parameter $<1$. The last term in Eq.~(\ref{ham_arr}) describes the coherent tunneling of Cooper-pairs between neighboring islands. Here, the operators $n_i,\phi_j$ are canonically conjugate satisfying $[n_i,e^{\pm i\phi_j}] = \pm e^{\pm i\phi_j}\delta_{ij}$. The Hamiltonian can be viewed as a Bose-Hubbard model with nearest neighbor interaction, in the limit of very high-occupancy at each site and zero gate voltage~\cite{Fisher1989} (see~\ref{phase_diag} for more details) \footnote{In general, there is a term $E_g\sum_i n_i$ in the lattice Hamiltonian, which corresponds to `chemical potential' in the Bose-Hubbard language and $E_g$ is the gate voltage, see~\ref{phase_diag}. In this section, this term is set to zero by choosing the `chemical potential' appropriately.}.

The boundary conditions have simple physical interpretations for the circuit model. The Dirichlet boundary condition at an end corresponds to a fixed superconducting phase at that end. As a result, there is no voltage drop at the: $V\sim \partial_t\phi = 0$ and there is short-circuit at the boundaries~\cite{Devoret_1997, Roy2016, Roy2018}. This can be achieved by adding a Josephson junction with a very large junction energy compared to its charging energy at the boundary. The Neumann boundary condition at and end corresponds to leaving the end open. As a result, no current can flow $I\sim \partial_x\phi=0$~\cite{Roy2016, Roy2018}. In the DMRG simulations, we implement the boundary conditions by appropriately choosing the boundary interaction terms. 

\subsubsection{DMRG results:}
\label{dmrg_res}
The DMRG simulations were performed using the TeNPy package~\cite{Hauschild2018}. The local Hilbert space on each island was truncated to 9: $n_i  = -4,-3,\ldots,3,4$. For definiteness, we chose $\delta=0.2$ and $\langle n_i\rangle =0$ by choosing an appropriate ground state sector. Furthermore, we chose a maximum bond-dimension of $500$ to keep the errors in truncation below $10^{-9}$. Here, we provide only the results relevant for the free, compactified boson CFT (the details of the phase-diagram can be found in \ref{phase_diag}). We perform DMRG simulations for both  open Neumann and Dirichlet  boundary conditions at the ends of the lattice. 

\begin{figure}
\centering
\includegraphics[width = 0.9\textwidth]{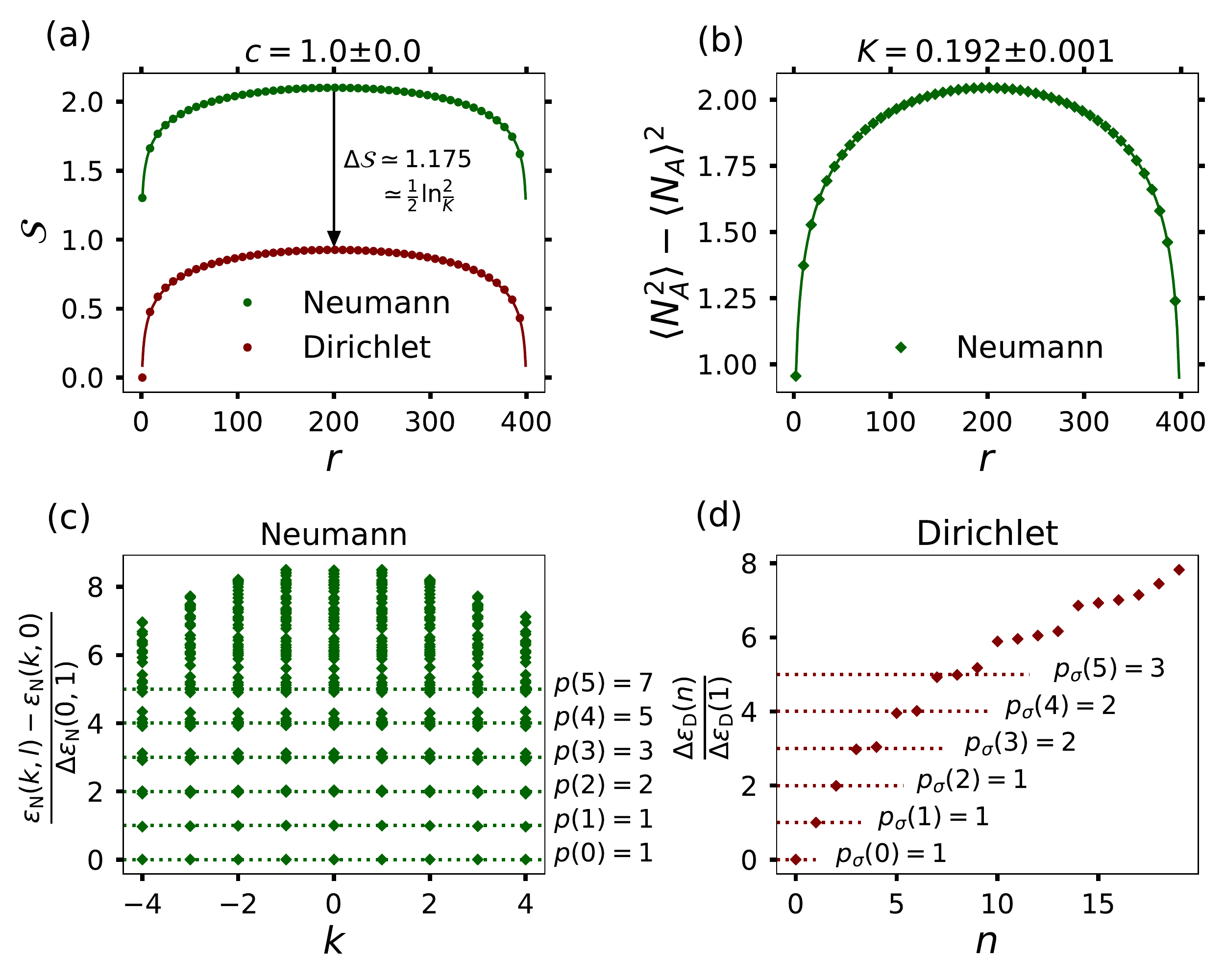}
\caption{\label{lutt_cent_charge} DMRG results for the Neumann (dark green) and Dirichlet (maroon) boundary conditions. The parameters are chosen to be $E_{J}/E_{C_0} = 8$, $\delta = 0.2$ for the quantum circuit lattice model of Eq.~(\ref{ham_arr}) so as to be in the free, compactified boson phase. The  system size $L_0 = 400$. (a) Entanglement entropy, ${\cal S}(r,L_0)$, as a function of subsystem size ($r$) for Neumann and Dirichlet boundary conditions. The central charge is extracted from the Neumann case and is obtained to be $\simeq1$. As the boundary condition is changed from Neumann to Dirichlet, the entanglement entropy changes by $\simeq1.175\simeq[\ln (2/K)]/2$, where $K$ is the Luttinger parameter.
  (b) The Luttinger parameter extracted from the variation of the particle number fluctuations within subsystem A as a function of the $r$ [see Eq.~(\ref{deltaNA})] is given by $K\simeq0.192$. 
  (c) Rescaled entanglement spectrum for Neumann boundary conditions. The entanglement energies $\varepsilon_N(k,l)$ are plotted as a function of $k$. The latter determines the dimension of the primary field of the boundary CFT, which are given by $Kk^2/2$. For a given $k$, from Eq.~(\ref{NN_disp}), the y-axis yields the level of the descendant field, indexed by $l$. The CFT predictions for the magnitudes of the eigenvalues are indicated by dashed lines, while the expected degeneracies are given by the integer partitioning of $l$, denoted by $p(l)$.
  (d) Rescaled entanglement spectrum for Dirichlet boundary conditions. The entanglement energies $\varepsilon_D(n)$ are plotted as a function of $n$ [see Eq.~(\ref{DN_disp})]. The CFT predictions for the magnitude of the rescaled entanglement energies and their corresponding degeneracies are indicated by the dashed lines and the $p_\sigma(n)$ respectively [see Eqs.~(\ref{psigma}, \ref{DN_disp})]. 
}
\end{figure}

First, we obtain the two main characteristics of the free, compactified boson CFT: the central charge and the compactification radius or equivalently the Luttinger parameter. The first is obtained by computing the entanglement entropy for the subsystem A as a function of the subsystem size $r$ [see Eq.~(\ref{ent_form})]. The results are shown in Fig.~\ref{lutt_cent_charge}(a) for both Neumann (dark green) and Dirichlet (maroon) boundary conditions for a system size $L_0 = 400$. The central charge is extracted from the data for the Neumann boundary conditions and is obtained to be $\simeq1$ as expected. As the boundary condition is changed, the entanglement entropy changes by the expected amount of $\ln(2/K)/2$, where $K$ is the Luttinger parameter. This the contribution from the boundary entropy~\cite{Affleck1991, Fendley1994} as computed in Eq.~(\ref{eND_b}) [note the extra factor of $1/2\pi$ in the latter equation due to conventions chosen in Eq.~(\ref{ent_ham})].
To obtain the Luttinger parameter, we compute the particle number fluctuations within the subsystem as a function of $r$. This yields the Luttinger parameter through the following relation~\cite{Rachel2012}
\begin{eqnarray}
  \label{deltaNA}
(\Delta N_A)^2 &= \langle N_A^2 \rangle - \langle N_A\rangle^2= \frac{1}{2\pi^2 K}\ln \Big(\frac{2L_0}{\pi a}\sin\frac{\pi r }{L_0}\Big) + \Delta N_0,
\end{eqnarray}
where $\Delta N_0$ is some non-universal contribution. The result for Neumann boundary condition is shown in Fig.~\ref{lutt_cent_charge}(b), where the Luttinger parameter is obtained to be $K\simeq0.192$, with error bars occurring in the third decimal place. Here, we choose to extract the Luttinger parameter using particle number fluctuations as opposed to real-space correlations since the simulations are done for finite size systems and we wish to avoid boundary effects. The Luttinger parameter extracted was checked to agree with that extracted from real-space correlation function using infinite DMRG within error-bars (see~\ref{phase_diag} for such computations). 
Next, we compute the entanglement spectrum for the Neumann and Dirichlet boundary conditions using DMRG. The results are shown in Fig.~\ref{lutt_cent_charge}(c) and (d) respectively.  The entanglement spectrum is computed by partitioning the system in half: $r = L_0/2$. To relate to Eqs.~(\ref{NN_disp}, \ref{DN_disp}), we plot the rescaled entanglement energies. For the Neumann case, we plot $[\varepsilon_N(k,l) - \varepsilon_N(k,0)]/\Delta\varepsilon_N(0,0)$ [Fig.~\ref{lutt_cent_charge}(c)] as a function of $k$. The latter determine the dimension of the primary fields, which are $Kk^2/2$. From Eq.~(\ref{NN_disp}), the y-axis is the level of the descendant field, indexed by $l$. The CFT predictions for each level is indicated by dashed dark green lines and the corresponding degeneracies are given by the integer partitioning of $l$, denoted by $p(l)$. Fig.~\ref{lutt_cent_charge}(d) shows the rescaled entanglement energies $\Delta\varepsilon_D(n)/\Delta\varepsilon_D(1)$ for the Dirichlet boundary conditions. The corresponding CFT predictions for the entanglement energies and the corresponding degeneracies are indicated by dashed maroon lines and $p_\sigma(n)$ respectively.  

\begin{figure}
\centering
\includegraphics[width = 0.9\textwidth]{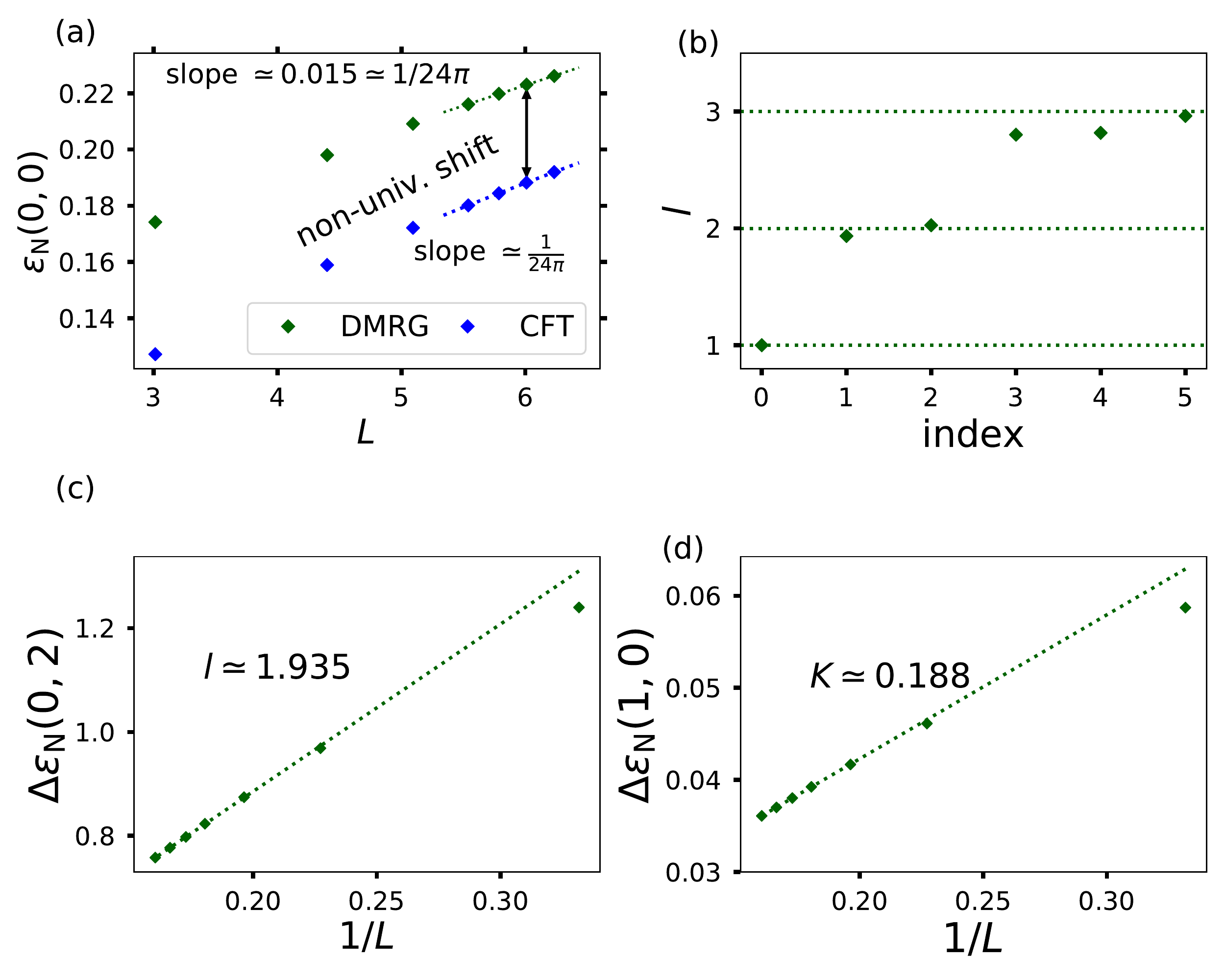}
  \caption{\label{e0_varn} (a) Comparison of DMRG vs CFT results for the variation of $\varepsilon_N(0,0)$ with $L = \ln(2L_0/\pi)$. Asymptotically, for large $L$, $\varepsilon_N(0,0)$ obtained with DMRG (in maroon) shows a linear dependence with a slope close to the expected value of $1/24\pi$ [see Eq.~(\ref{e_0})]. The CFT prediction is plotted in blue. There is a non-universal constant shift between the CFT and the DMRG results. The latter can be absorbed in the speed of sound of the boundary CFT that is set to 1 in the computations of~\ref{bcft}.
  (b) Extracted degeneracies at the different values of $l$ from computing the variation of $\Delta\varepsilon_N(0,l)$ vs $L$ for the first six entanglement eigenvalues, confirming the CFT predictions. Higher energy levels also show the clustering predicted by the CFT computations, but the data quality diminishes due to finite truncation errors and are not shown. 
  (c) Variation of $\Delta\varepsilon_N(0,2)$ vs $1/L$ [see Eq.~(\ref{NN_disp})]. As system size increases (small values of $1/L$), we get the expected linear dependence. From the normalized slope (see main text for the details on the normalization), we extract the value of $l$ to $\simeq1.935$ (which is close to the expected value: 2).  (d) Variation of $\Delta\varepsilon_N(1,0)$ vs $1/L$. From Eq.~(\ref{NN_disp}), the dependence is of the form $\sim \pi K/2L$. From the linear dependence, after normalization, we extract the Luttinger parameter to be $K\simeq0.188$, which is close to what was obtained using particle number fluctuations in Fig.~\ref{lutt_cent_charge}.
  }
\end{figure}

Now, we analyze the finite size dependence of the lowest few entanglement energies for the Neumann case (similar analysis was done for the Dirichlet case and are not shown for brevity). First, consider the lowest entanglement energy $\varepsilon_N(0,0)$. We again choose the subsystem size $r=L_0/2$ and plot the variation with respect to $L = \ln(2L_0/\pi)$ [see Eq.~(\ref{ldef})]. The variation of the lowest entanglement energy with $L$ is shown in maroon in Fig.~\ref{e0_varn}. As the subsystem size is increased, $\varepsilon_N(0,0)$ shows the expected linear dependence [see Eq.~(\ref{e_0})] with slope $0.015\sim 1/24\pi$. The corresponding values of $\varepsilon_N(0,0)$ obtained from the CFT predictions of Eq.~(\ref{e_0}) is plotted in blue. In the limit of large $L$, we see that there is a non-universal constant shift between the CFT prediction and the DMRG result. This non-universal shift is generically present in the equality given in Eq.~(\ref{ent_ham_formula}). This is because this shift can be absorbed by rescaling the speed of sound of the boundary CFT (the latter is set to unity in the computations of~\ref{bcft}). 
Finally, we analyze the variation of the higher entanglement energy levels, $\varepsilon_N(k,l)$, in each Virasoro tower [see Fig.~\ref{lutt_cent_charge}(c)] as a function of system size. From  Eq.~(\ref{NN_disp}), for the Virasoro tower built on top each primary field indexed by $k$, $\varepsilon_N(k,l) - \varepsilon_N(k,0)=\pi l/L$, where $L$ is defined in Eq.~(\ref{ldef}). Thus, from this dependence we can get the dimension of the descendants indexed by $l$, which should exhibit the degeneracy given by $p(l)$. We do this for $k=0$ (similar results were obtained for other $k$-s and are not shown for brevity). On the other hand, the variation of $\Delta\varepsilon_N(k,0)$ vs $1/L$ yields the Luttinger parameter $K$ [see Eq.~(\ref{NN_disp})]. We do this analysis for $k=1$. In order to remove the non-universal effects due to the lattice, we needed to normalize the obtained values of $l$ and $K$ by a non-universal parameter given by the slope of the variation of $\Delta\varepsilon_N(0,1)$ vs $l/L$. The latter is not the expected value of $\pi$ as predicted by the CFT computations. This is because  the values of $L=\ln(2L_0/\pi)$ are quite small  despite the overall system size, $L_0$, being up to 800. Fig.~\ref{e0_varn}(b) shows the expected degeneracies for $l=1,2,3$. Fig.~\ref{e0_varn}(c) shows the variation of $\Delta\varepsilon_N(0,2)$ vs $1/L$, which after the normalization process described above yields $l\simeq 1.935$ which is close to the expected value of 2. Finally, Fig.~\ref{e0_varn}(d) shows the variation of $\Delta\varepsilon_N(1,0)$ vs $1/L$, which yields a Luttinger parameter of $K\simeq0.188$ which is close to the value obtained earlier using particle number fluctuations (see Fig.~\ref{lutt_cent_charge}). We believe the discrepancy between the obtained values of the Luttinger parameter (here from finite-size scaling and that from particle number fluctuations) arises due to the scaling with $L=\ln(2L_0/\pi)$ which is relatively small despite the overall system size, $L_0$, being up to 800.

\section{Summary and Perspectives}
\label{concl}
In this work, we have analytically computed the spectra of the entanglement/modular Hamiltonian of the Ising and the free, compactified boson  CFTs in terms of the spectra of corresponding boundary CFTs. The boundary CFTs were analyzed by computing the corresponding partition functions using the relevant Ishibashi states. We compared the analytical predictions for the continuum theory by numerically analyzing corresponding lattice regularized models. In contrast to traditional approaches of using the XXZ chain as a lattice-regularization for the compactified boson CFT, in this work, we analyzed a quantum circuit model using an array of Josephson junctions. While the quantum circuit lattice model is non-integrable to the best of our knowledge, in the long-distance limit, for appropriate choice of parameters, it gives rise to the relevant CFT. We start with lattice degrees of freedom that are directly the discretized, compact bosonic fields being simulated. We investigate the lattice model with DMRG. We showed that the CFT and the DMRG predictions are compatible with each other, up to non-universal renormalization of the entanglement spectrum including overall shifts and scale factors. These non-universal effects can be absorbed by rescaling the velocity of sound in the boundary CFT computations. 

Exact computation of entanglement Hamiltonians can also be performed in the case when there is a bulk perturbation to the CFT. In particular, consider the case when the perturbation preserves a subset of the infinite set of integrals of motion of a CFT, {\it i.e.,} the resultant is an integrable quantum field theory~\cite{Zamolodchikov1989}. In this case, the problem reduces to the computation of properties of integrable, perturbed, boundary-interacting CFTs~\cite{Cho2017}. In many cases, this computation is also analytically tractable. As a concrete example, consider the case of the quantum sine-Gordon model with Neumann boundary conditions at the ends. Then, the entanglement spectrum of the quantum sine-Gordon model is given by the spectrum of the boundary sine-Gordon model~\cite{Ghoshal1994}. The boundary sine-Gordon model has Neumann boundary condition at one end (this is inherited from the original model with bulk perturbation, see Fig.~\ref{1D_system}) and a cosine potential at the other end (this boundary condition arises from the entanglement cut). The spectrum and the boundary S-matrices for this model are well-known. We aim to analyze the problem in the context of the entanglement Hamiltonian of the quantum sine-Gordon model in the near future. 

Finally, we note that the entanglement Hamiltonians provide a fruitful method to investigate the physical spectrum of perturbed CFTs using DMRG. In general, with the latter numerical tool, the computational complexity to evaluate the physical spectrum beyond the first few levels grows rapidly. However, the entanglement spectrum, which is contained in the Schmidt decomposition of the system at various bipartitionings, is the key ingredient of the DMRG analysis and can be evaluated with much more accuracy more easily. 

\section{Acknowledgments}
Discussions with Johannes Hauschild are gratefully acknowledged. FP and AR are funded by the European Research Council (ERC) under the European Unions Horizon 2020 research and innovation program (grant agreement No. 771537). FP acknowledges the support of the DFG Research Unit FOR 1807 through grants no. PO 1370/2-1, TRR80, and the Deutsche Forschungsgemeinschaft (DFG, German Research Foundation) under Germany’s Excellence Strategy EXC-2111-390814868. HS was supported in part by the Advanced ERC NuQFT. 
\appendix
\section{Boundary CFT of the free compactified boson}
\label{bcft}
\begin{figure}
\includegraphics[width = 0.55\textwidth]{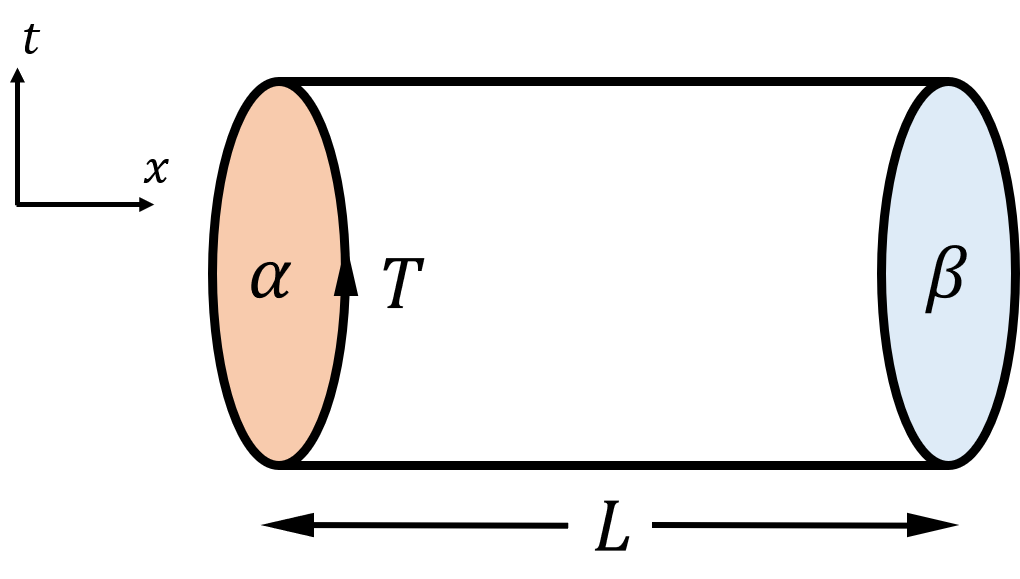}
\caption{\label{bcft_pic} Schematic of a boundary CFT over spatial interval $L$ with boundary conditions at $x=0(L)$ given by $\alpha(\beta)$. The extent in the imaginary time direction is given by $T$. }
\end{figure}

In this section, we provide a derivation of the boundary states and the partition function of the free compactified boson theory using the general formalism of boundary CFTs~\cite{Cardy1989, diFrancesco1997}, extending the calculation done in Ref.~\cite{Oshikawa1997}. We start with the Euclidean action for the free compactified boson CFT on a finite length interval $L$ given by
\begin{equation}
S = \frac{g}{2}\int_0^T dt\int_0^L dx\Big[(\partial_t\phi)^2 + (\partial_x\phi)^2\Big].
\end{equation}
Periodic boundary condition is imposed in the imaginary time direction, the length of which is given by $T$. The boundary condition at $x=0(L)$ is given by $\alpha(\beta)$ (see Fig.~\ref{bcft_pic}). 
The compactification radius is $R$, so that $\phi(x,t)$ may be identified by $\phi(x,t) + 2\pi R$. The partition function with boundary condition $(\alpha, \beta)$ is given by
\begin{equation}
Z_{\alpha\beta}(q) \equiv {\rm Tr}\ e^{-T H_{\alpha\beta}} = {\rm Tr}\ q^{LH_{\alpha\beta}/\pi},
\end{equation}
where $q \equiv e^{2\pi i \tau} = e^{-\pi T/L}$, where the modular parameter $\tau = iT/2L$. Furthermore, $H_{\alpha\beta}$ is the Hamiltonian of the boundary CFT with boundary conditions $\alpha,\beta$. The spectrum of $H_{\alpha,\beta}$ falls into irreducible representations of the Virasoro algebra~\cite{diFrancesco1997}
\begin{equation}
\label{qpic}
Z_{\alpha\beta}(q) = \sum_i n^i_{\alpha\beta} \chi_i(q),
\end{equation}
 where the Virasoro character $\chi_i(q)$ is given by \begin{equation}
\chi_i(q) = q^{-1/24}{\rm Tr}_i\ q^{L_0},
\end{equation}
where we have used the fact that the central charge is 1 and $L_0$ is the relevant Virasoro generator. The partition function can equally well be written in the modular transformed picture $\tau\rightarrow -1/\tau$. In this picture, 
\begin{equation}
Z_{\alpha\beta}(q) = \langle \alpha| e^{-L \bar{H}}|\beta\rangle,
\end{equation}
where $\bar{H}$ is the Hamiltonian of the interval $T$ with periodic boundary conditions and $|\alpha,\beta\rangle$ are the corresponding boundary states. Below, we compute these boundary states and evaluate this partition function. It is convenient to map the cylinder to the plane using $z = e^{T(t-ix)/2\pi}$. Then, 
\begin{equation}
\bar{H} = \frac{2\pi}{T}\Big(L_0^z + \bar{L}_0^z -\frac{1}{12}\Big),
\end{equation}
where we have labeled the Virasoro generators in the $z$-plane by a superscript to emphasize the fact that on the $z$-plane, the holomorphic and the anti-holomorphic components propagate separately, 
which is different from those defined in Eq.~(\ref{qpic}). This leads to 
\begin{equation}
Z_{\alpha, \beta}(q) = \langle \alpha |(\tilde{q}^{1/2})^{L_0^z + \bar{L}^z_0 -1/12}|\beta\rangle, \ \tilde{q} \equiv e^{-4\pi L/T}.
\end{equation}
Expanding in normal modes (see Chap. 6.3.5 of Ref.~\cite{diFrancesco1997}), we can write
\begin{eqnarray}
\phi(z,\bar{z}) &= \phi_0 -i\Big(\frac{n}{4\pi gR}+\frac{mR}{2}\Big)\ln z\nonumber\\&\quad -i \Big(\frac{n}{4\pi gR}-\frac{mR}{2}\Big)\ln \bar{z}\nonumber\\&\quad + \frac{i}{\sqrt{4\pi g}}\sum_{k\neq0}\frac{a_k}{k}z^{-k}+ \frac{i}{\sqrt{4\pi g}}\sum_{k\neq0}\frac{\bar{a}_k}{k}\bar{z}^{-k},
\end{eqnarray}
where $m$ is the winding number and $n$ is the quantization of the zero-mode momenta. Here, we have defined
\begin{eqnarray}
a_k &= -i\sqrt{k}\tilde{a}_k, \ k>0,\nonumber\\& =i\sqrt{-k}\tilde{a}_{-k}^\dagger,\ k<0,\\\bar{a}_k &= -i\sqrt{k}\tilde{a}_{-k}, \ k>0,\\& =i\sqrt{-k}\tilde{a}_k^\dagger,\ k<0,
\end{eqnarray}
where $\tilde{a}_k$ are the original bosonic operators satisfying $[\tilde{a}_k,\tilde{a}_l] = 0$, $[\tilde{a}_k,\tilde{a}_l^\dagger]=\delta_{kl}$. 
Straightforward computations show that 
\begin{eqnarray}
L_k &= \frac{1}{2}\sum_l :a_la_{k-l}:,\ \bar{L}_k = \frac{1}{2}\sum_l :\bar{a}_l\bar{a}_{k-l}:,
\end{eqnarray}
where we have defined 
\begin{eqnarray}
a_0, \bar{a}_0 = \sqrt{4\pi g}\Big(\frac{n}{4\pi gR}\pm\frac{mR}{2}\Big).
\end{eqnarray}
The Hamiltonian, $\bar{H}$, is given by 
\begin{eqnarray}
\bar{H} &= \frac{2\pi}{T}\Big[\Big(\frac{n^2}{4\pi gR^2} + m^2R^2\pi g\Big)-\frac{1}{12}\nonumber\\&\qquad + \sum_{k>0}k\big(a_{-k}a_k + \bar{a}_{-k}\bar{a}_k\big)\Big]
\end{eqnarray}

In the boundary CFT, a boundary state, $|B\rangle$, satisfies the Ishibashi condition:~\cite{Ishibashi1988}
\begin{equation}
\label{ishibashi}
(L_k - \bar{L}_{-k})|B\rangle = 0, \ \forall k.
\end{equation}
For the compactified boson, Ishibashi conditions for the the Dirichlet (D) and Neumann (N) boundary conditions are satisfied if:~\cite{Oshikawa1997}
\begin{eqnarray}
\Big[a_k -(+) \bar{a}_{-k}\Big]|D(N)\rangle = 0,
\end{eqnarray}
where we have denoted the corresponding boundary states by $|D(N)\rangle$. The boundary states can be constructed by applying appropriate operators on the vacua labeled by the zero-mode indices $(n,m)$. Since the Dirichlet (Neumann) states have to satisfy the Ishibashi condition [Eq.~(\ref{ishibashi})] for $k=0$, this implies the Dirichlet (Neumann) states are built on vacua labeled by $|n,0\rangle$ ($|0,m\rangle$). By using the bosonic commutation relations, it is easy to see that 
\begin{eqnarray}
|D\rangle &= \sum_n c_n {\rm{exp}}\Big[-\sum_{k>0}\tilde{a}_{-k}^\dagger \tilde{a}_{k}^\dagger\Big]|n,0\rangle\\|N\rangle &= \sum_m d_m {\rm{exp}}\Big[+\sum_{k>0}\tilde{a}_{-k}^\dagger \tilde{a}_{k}^\dagger\Big]|0,m\rangle,
\end{eqnarray}
where $c_n,d_m$ are coefficients that need to be determined. The overall normalization of the two states are fixed by imposing the Cardy consistency condition that exactly one dimension zero character.~\cite{Cardy1989} We directly provide the boundary states following Ref.~\cite{Oshikawa1997} and check for the consistency afterwards~\footnote{We correct some errors in the cited reference.}. The Dirichlet and Neumann boundary states are given by 
\begin{eqnarray}
|D(\phi_0)\rangle &= \frac{1}{\sqrt{2R\sqrt{\pi g}}}\sum_{n}e^{-\frac{in\phi_0}{R\sqrt{\pi g}}}{\rm{exp}}\Big[-\sum_{k>0}\tilde{a}_{-k}^\dagger \tilde{a}_{k}^\dagger\Big]|n,0\rangle,\\
|N(\tilde{\phi}_0)\rangle &= \sqrt{R\sqrt{\pi g}}\sum_{m}e^{-\frac{im\tilde{\phi}_0R}{2\sqrt{\pi g}}}{\rm{exp}}\Big[+\sum_{k>0}\tilde{a}_{-k}^\dagger \tilde{a}_{k}^\dagger\Big]|0,m\rangle,
\end{eqnarray}
where we have also used the duality between Neumann and Dirichlet boundary conditions: $R\leftrightarrow2/R$ and $\tilde{\phi}_0$ is the field dual to $\phi_0$. It is easy to check that under Dirichlet (Neumann) boundary conditions, the field $\phi(\tilde\phi)$ is pinned to the value $\phi_0(\tilde\phi_0)$ at the boundary. Next, we compute the partition functions for different combinations of boundary conditions. 

\subsection{Dirichlet-Dirichlet boundary condition}
\label{sec_DD}
Consider the case when Dirichlet boundary conditions are imposed both at $x=0$ and $x=L$. Denote the corresponding boundary states by $|D(\phi_0)\rangle, |D(\phi_0')\rangle$. Then,
\begin{eqnarray}
Z_{\rm{DD}}(q) &= \langle D(\phi_0)|e^{-L\bar{H}}|D(\phi_0')\rangle\nonumber\\&=\frac{1}{2R\sqrt{\pi g}}\sum_n e^{\frac{in\Delta\phi_0}{R\sqrt{\pi g}}}e^{-\frac{2\pi L}{T}\big(\frac{n^2}{4\pi gR^2}-\frac{1}{12}\big)}\prod_{k>0}\frac{1}{1-e^{-\frac{4\pi Lk}{T}}}\nonumber\\& = \frac{1}{2R\sqrt{\pi g}}\frac{1}{\eta(\tilde{q})}\sum_n  e^{\frac{in\Delta\phi_0}{R\sqrt{\pi g}}}\tilde{q}^{\frac{n^2}{8\pi gR^2}},
\end{eqnarray}
where $\Delta\phi_0 = \phi_0 - \phi_0'$ and $\eta(q)$ is the Dedekind function defined in Eq.~(\ref{Dedekind}). In order to express the result in terms of $q$, we use the following relation~\cite{Apostol1997}: 
\begin{equation}
\eta(\tilde{q}) =\sqrt{\frac{T}{2L}}\eta(q)
\end{equation}
and the Poisson summation formula: 
\begin{equation}
\sum_n e^{-\pi a n^2 + b n} = \frac{1}{\sqrt{a}}\sum_k e^{-\frac{\pi}{a}\big(k + \frac{b}{2i\pi}\big)^2}. 
\end{equation}
After simple manipulations, we arrive at 
\begin{equation}
Z_{\rm{DD}}(q) = \frac{1}{\eta(q)}\sum_k \big(q^{2\pi g R^2}\big)^{\big(k + \frac{\Delta\phi_0}{2\pi R\sqrt{\pi g}}\big)^2}.
\end{equation}
Now, consider the case when $\Delta\phi_0 = 0$. For this case,
\begin{eqnarray}
Z_{\rm{DD}}(q) &= \frac{1}{\eta(q)}\sum_k q^{2\pi g R^2k^2}\nonumber\\& = \sum_h n^h_{\rm{DD}}\chi_h(q)= \sum_h n^h_{\rm{DD}}\frac{q^h}{\eta(q)},
\end{eqnarray}
where we have used Eq.~(\ref{qpic}) and the definition of $\eta(q)$. From the above equation, we see that $n^{h=0}_{\rm{DD}}=1$, as expected from the Cardy consistency relations~\cite{Cardy1989}. The other primary fields have dimensions $2\pi g R^2k$ where $k\neq 0$. 
Using the explicit expression of $\eta(q)$:
\begin{eqnarray}
\frac{1}{\eta(q)}\equiv q^{-1/24}\frac{1}{\varphi(q)} \equiv \sum_{l\geq0}p(l)q^{-1/24+l},
\end{eqnarray}
where $p(l)$ is the number of ways to partition the integer $l$, we arrive at 
\begin{eqnarray}
Z_{\rm{DD}}(q) &= {\rm Tr}\ q^{LH_{\alpha,\beta}/\pi} \nonumber\\& = \sum_k \sum_{l\geq0}p(l)q^{-1/24 + 2\pi g R^2 k^2 + l}.
\end{eqnarray}
Thus, the boundary CFT has the spectrum given by 
\begin{equation}
\epsilon_{\rm{DD}}(k,l) = \frac{\pi}{L}\Big(-\frac{1}{24} + 2\pi g R^2 k^2 + l\Big), \ k\in\mathbb{Z}, l\geq0,
\end{equation}
with degeneracy $p(l)$. The spectrum can be thought of as composed of towers on each primary field with dimension $2\pi gR^2k$ and the descendants being indexed by $l$. For each tower, the descendants at level $l$ have degeneracy $p(l)$. The compactification radius is related to the Luttinger parameter $K$ by
\begin{equation}\label{comp}
R = \frac{1}{\sqrt{\pi K}}.
\end{equation}
Furthermore, it is convenient to choose $g=1$ as the normalization of the free-boson action. Then, the spectrum is given by
\begin{equation}
\label{DD_spec}
\epsilon_{\rm{DD}}(k,l) = \frac{\pi}{L}\Big(-\frac{1}{24} + \frac{2}{K}k^2 + l\Big), \ k\in\mathbb{Z}, l\geq0.
\end{equation}

\subsection{Neumann-Neumann boundary conditions}
\label{sec_NN}
Now, consider the case when Neumann boundary conditions is imposed on both ends $x=0,L$. Denote the boundary states by $|N(\tilde{\phi}_0)\rangle, |N(\tilde\phi_0')\rangle$. Similar calculation as in the Dirichlet case leads to 
\begin{equation}
Z_{\rm{NN}}(q) = \frac{R\sqrt{\pi g}}{\eta(\tilde{q})}\sum_m e^{\frac{imR\Delta\tilde\phi_0}{2\sqrt{\pi g}}}\tilde{q}^{\frac{m^2R^2\pi g}{2}},
\end{equation}
which can be rewritten in terms of $q$ as 
\begin{equation}
Z_{\rm{NN}}(q) = \frac{1}{\eta(q)}\sum_m \big(q^{\frac{1}{2R^2\pi g}}\big)^{\big(m + \frac{R\Delta\tilde\phi_0}{4\pi\sqrt{\pi g}}\big)^2}.
\end{equation}
For $\Delta\tilde\phi_0=0$, using Eq.~(\ref{comp}), we get 
\begin{eqnarray}
Z_{\rm{NN}}(q) &= \frac{1}{\eta(q)}\sum_k q^{\frac{K}{2 g}k^2} = \sum_h n^h_{\rm{NN}}\frac{q^h}{\eta(q)}\\&=\sum_k\sum_{l\geq0}p(l)q^{-\frac{1}{24} +\frac{Kk^2}{2g} + l}
\end{eqnarray}
We see that the Cardy consistency condition is again satisfied: $n^{h=0}_{\rm{NN}} = 1$ and the spectrum of the boundary CFT is given by
\begin{equation}
\label{NN_spec}
\epsilon_{\rm{NN}}(k,l) = \frac{\pi}{L}\Big(-\frac{1}{24} + \frac{K}{2}k^2 + l\Big), \ k\in\mathbb{Z}, l\geq0
\end{equation}
with degeneracy $p(l)$. 

\subsection{Dirichlet-Neumann boundary conditions}
\label{sec_ND}
Finally, we consider the case when Neumann boundary condition is imposed on the end $x=0$ and Dirichlet on the end $x=L$. In this case, the partition function is computed by evaluating $\langle D(\phi_0)|e^{-L\bar{H}}|N(\tilde{\phi}_0)\rangle$. The rest of the steps are identical as in the previous cases and lead to 
\begin{eqnarray}
\label{part_fun_ND}
  Z_{\rm DN}(q) & = &\frac{1}{\sqrt{2}\eta(\tilde{q})}\sum_n  (-1)^n \tilde{q}^{n^2}\\&=& \frac{1}{2\eta(q)}\sum_k q^{\frac{1}{4}\big(k+\frac{1}{2}\big)^2}\\&=&\frac{q^{1/48}}{2\varphi(q)}\sum_k q^{\frac{k^2 + k}{4}} = q^{1/48}\sum_{n\geq0}q^{n/2}p_\sigma(n)
  %\\&=\sum_{k\geq0} \sum_{l\geq0}p(l)q^{-\frac{1}{24} + \frac{1}{4}\big(k+\frac{1}{2}\big)^2 + l}.
\end{eqnarray}
where $p_\sigma(n)$ are the degeneracy factors that occur in the Ising CFT with Dirichlet boundary conditions, given in Eq.~(\ref{psigma}). 
Thus, the spectrum in this case is given by 
\begin{equation}
\label{ND_spec}
  \epsilon_{\rm{DN}}(n) = \frac{\pi}{L}\Big(\frac{1}{48}+\frac{n}{2}\Big),
\end{equation}
where $n\geq0$ and the degeneracies are given by $p_\sigma(n)$ at level $n$. 

\section{Phase-diagram of the quantum circuit model}
\label{phase_diag}
\begin{figure}
\centering
\includegraphics[width = \textwidth]{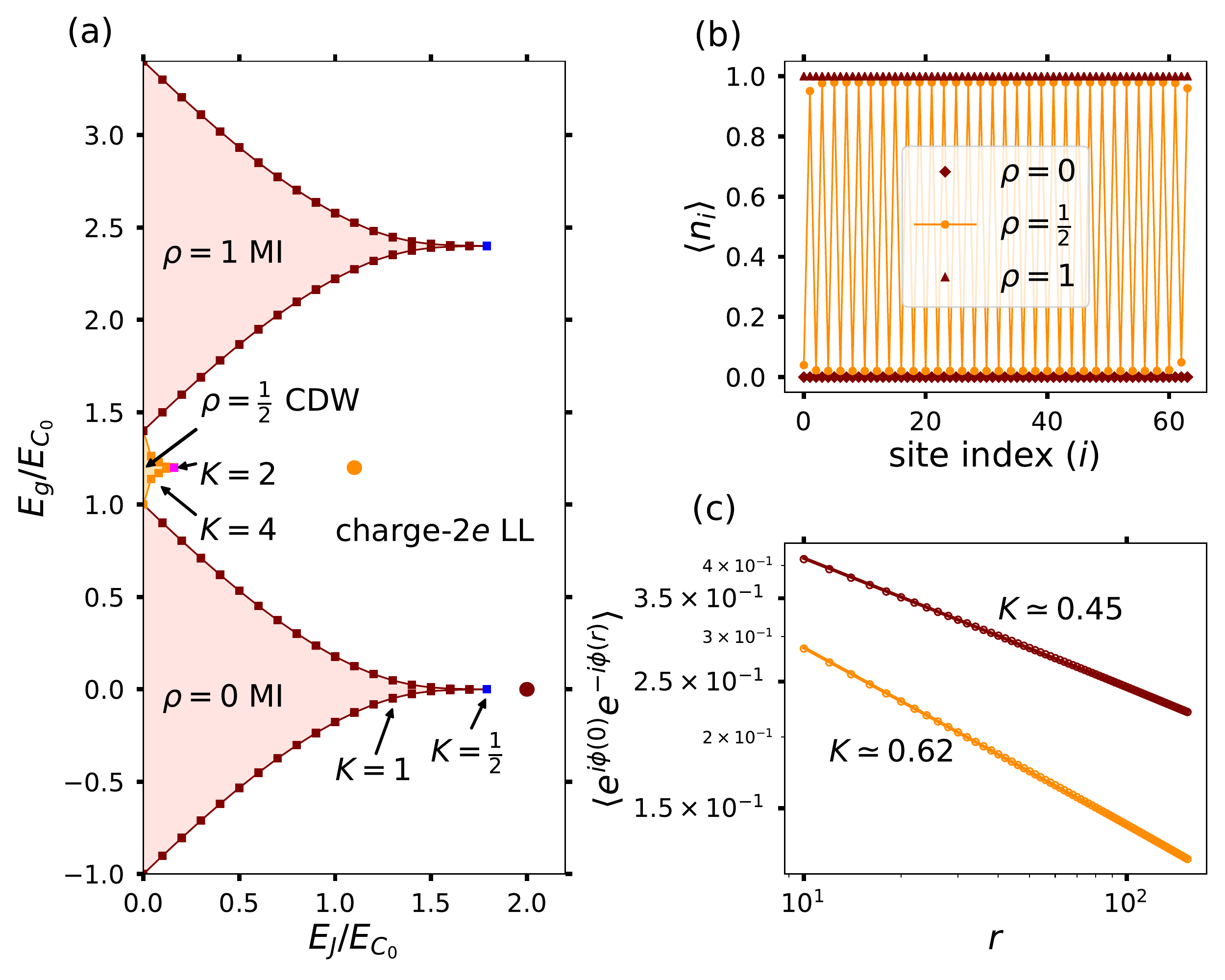}
  \caption{\label{lobes} We chose $\delta = 0.2$ [see Eq.~(\ref{ham_arr_1})]. (a) Phase-diagram of the quantum circuit model as a function of $E_J/E_{C_0}$ and $E_g/E_{C_0}$. Within the maroon lobes, the system is in a Mott-insulating (MI) phase, where the density of Cooper-pairs ($\rho$) on each island is pinned to an integer. We show two such lobes with $\rho=0,1$. Between the two MI lobes, there is an orange lobe, where the system is in a charge-density-wave (CDW) phase, with $\rho=1/2$. In this phase, the average number of Cooper-pairs is pinned to an alternating Neel order: $01010\ldots$. Changing $E_J/E_{C_0}$ or $E_g/E_{C_0}$ causes the system to transition from the MI or the CDW phase into a charge-$2e$ Luttinger liquid (LL) phase, where the system is described by a free, compactified boson CFT, with the compactification radius determined by the Luttinger parameter ($K$). The transition from the tip of the lobes is of the type Kosterlitz-Thouless. At the transition from the MI (CDW) phase, indicated by the blue (magenta) square, $K=1/2(2)$. The transition from the sides of the lobes occur with a change in the Cooper-pair density. At the sides of the MI (CDW) lobes, indicated by the maroon (dark orange) squares, $K=1(4)$. (b) The average number of Cooper-pairs within the MI ($\rho = 0,1$) and the CDW ($\rho=1/2$) phases for a system size of 64. (c) The algebraic decay of correlations of the Cooper-pair creation and annihilation operators characteristic of the LL phase for two points, indicated by maroon and dark-orange circles in panel (a) obtained using infinite DMRG. For the chosen points, $K\simeq0.45$ and $K\simeq0.62$ respectively.}
\end{figure}

In this section, we analyze the phase-diagram of the quantum circuit model which, for appropriate choice of parameters, provides as a lattice-regularization of the free, compactified boson CFT. To that end, we start with the Hamiltonian of the model, given by (see Sec.~\ref{lattice_model})
\begin{eqnarray}
\label{ham_arr_1}
  H_{\rm{array}} &=& E_{C_0}\sum_{i=1}^Ln_i^2 + \delta E_{C_0}\sum_{i=1}^{L-1}n_in_{i+1} - E_g\sum_{i=1}^Ln_i\nonumber\\&&\quad - E_J\sum_{i=1}^{L-1}\cos(\phi_i - \phi_{i+1}),
\end{eqnarray}
where compared to Eq.~(\ref{ham_arr}), we have included an additional term corresponding to the gate-voltage at each superconducting island. This model can be viewed as a generalized Bose-Hubbard model in the limit of high-occupancy of the sites, where the role of bosons in played by Cooper-pairs. We work with the case when there is no disorder in the system. The charging energy terms (proportional to $E_{C_0}$) corresponds to onsite and nearest-neighbor repulsion terms, while the gate-voltage plays the role of the chemical potential. Finally, the Josephson tunneling term (proportional to $E_J$) gives rise to nearest-neighbor hopping. Note a crucial difference with the conventional Bose-Hubbard model. Here, $n_i$-s can be both positive and negative. Physically, $n_i$ corresponds to the {\it excess} number of Cooper-pairs on the $i^{\rm th}$ island. Thus, a negative $n_i$ would correspond to the removal of $|n_i|$ Cooper-pairs from the condensate on the $i^{\rm th}$ island. The phase-diagram of this model has been analyzed using perturbative analytical methods~\cite{Glazman1997, Fazio2001}. In what follows, we analyze the phase-diagram using DMRG generalizing the methods described in Ref.~\cite{Kuhner2000}. The local Hilbert space at each site was truncated at 9: $n_i=-4, -3, \ldots, 3,4$. Furthermore, we chose $\delta=0.2$.  

The phase-diagram obtained using DMRG is shown in Fig.~\ref{lobes}(a). Within the maroon lobes, the system is in a Mott-insulating (MI) phase. In this phase, the occupation of Cooper-pairs at each site is pinned to an integer, as shown in Fig.~\ref{lobes}(b). We only show the lobes for $\rho=0,1$ for brevity. Note that, in contrast to the conventional Bose-Hubbard model, the lobes extend in the negative $E_g/E_{C_0}$ regime because $n_i$-s can be negative. In addition to the MI lobes, the system can also be in a charge-density-wave (CDW) phase, that occurs in between two successive MI phases. This phase occurs due to the presence of nearest-neighbor repulsion in the model [$\delta\neq0$ in Eq.~(\ref{ham_arr_1})]. In this phase, the average densities of Cooper-pairs are half-integers, the case of $\rho=1/2$ is shown in dark-orange in Fig.~(\ref{lobes})(a,b), where the system shows an alternating Neel order for the Cooper-pair occupation, given by $010101\ldots$. Changing $E_J/E_{C_0}$ or $E_g/E_{C_0}$ causes the system to undergo a phase-transition into a charge-$2e$ Luttinger liquid (LL) phase, where the system is described by the free, compactified boson CFT. 
The compactification radius is determined by the Luttinger parameter $K$ [see Eq.~(\ref{comp})]. The latter determines the exponent of algebraic decay of correlations of the Cooper-pair creation and annihilation operators:
\begin{equation}
  \langle e^{i\phi(0)}e^{-i\phi(r)}\rangle\sim \frac{1}{|r|^{K/2}}
  \label{lutt-def}
\end{equation}
The transition through the tip of the lobes occurs at constant density of Cooper-pairs and is of the type Kosterlitz-Thouless. The location of the tip of the lobe was computed by locating the location where the $K$ crossed $1/2(2)$ for the MI(CDW) lobes. The phase-transition across the sides of the lobes occur with a change in the Cooper-pair density. The location of the sides of the lobe was computed by computing the values of $E_J$ for which the cost of adding/removing a particle is zero~\cite{Kuhner2000}. 
\begin{figure}
\centering
\includegraphics[width = 0.9\textwidth]{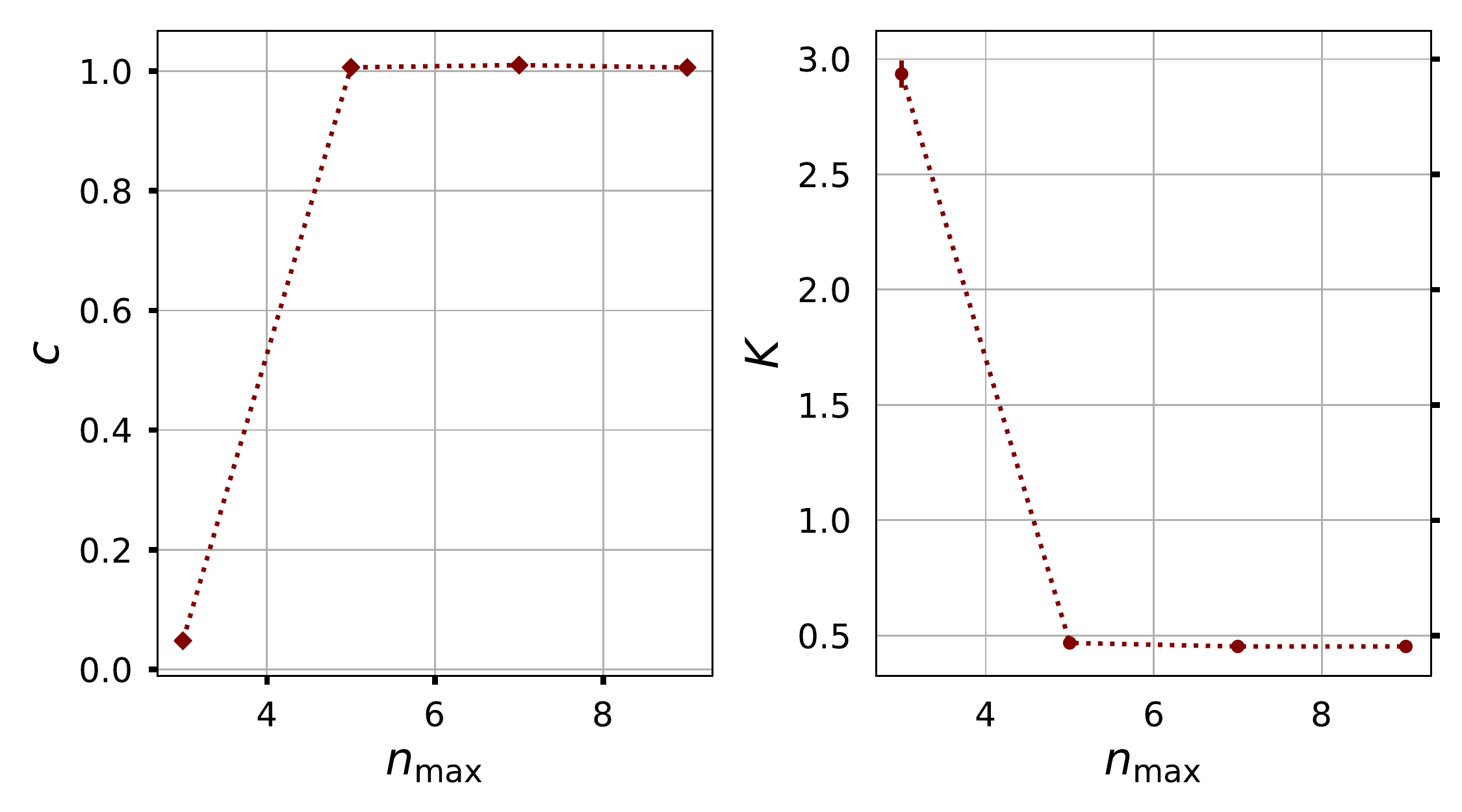}
\caption{\label{trunc_check} Variation of the central charge ($c$, left panel) and the Luttinger parameter ($K$, right panel) as the truncation cut-off ($n_{\rm{max}}$) is increased. As seen from both panels, the results are not affected for $n_{\rm{max}}\geq7$.}
\end{figure}
This characteristic decay in the LL phase and the obtained values of $K$ are shown in Fig.~\ref{lobes}(c) for two points, indicated by maroon and dark-orange circles in panel (a). The computation of the entanglement spectrum in the main-text was done for $E_g=0, E_J/E_{C_0}=8$. 

Finally, we present a numerical check that the truncation of the local Hilbert space at each site (chosen to be 9) is sufficient. To that end, we show the results for the maroon point in Fig.~\ref{lobes}, which lies along the constant density line $\langle n_i\rangle = 0$ (similar checks were performed for the other points). The free, compactified boson CFT has two characterizing features: the central charge ($c$) and the Luttinger parameter ($K$). The variation of these parameters with the truncation cut-off is presented in Fig.~\ref{trunc_check}. As is evident, the simulation results are not affected for $n_{\rm{max}}\geq7$. Note that this cut-off has to be appropriately chosen if the simulations are performed at higher densities.

\section*{References}
\bibliography{library_1}
\bibliographystyle{iopart-num}

\end{document}